\documentclass[acmsmall, screen, nonacm]{acmart}

\usepackage{hyperref}

\usepackage[T1]{fontenc}
\usepackage[utf8]{inputenc}
\usepackage{listings}
\usepackage{color}
\usepackage{natbib}
\usepackage{stmaryrd}
\usepackage{graphicx}
\usepackage{float}
\usepackage{placeins}
\usepackage{todonotes}

\definecolor{keywordcolor}{rgb}{0.7, 0.1, 0.1}   
\definecolor{commentcolor}{rgb}{0.4, 0.4, 0.4}   
\definecolor{symbolcolor}{rgb}{0.0, 0.1, 0.6}    
\definecolor{sortcolor}{rgb}{0.1, 0.5, 0.1}      
\definecolor{tacticcolor}{rgb}{0.0, 0.3, 0.6}    
\definecolor{attributecolor}{rgb}{0.7, 0.1, 0.1} 

\newcommand{\Nat}{\mathbb{N}}

\newcommand{\leanline}{\lstinline[language=lean]}
\newenvironment{leandisplay}
  {\par\addvspace{0.45\baselineskip}\begingroup\centering}
  {\par\endgroup\addvspace{0.3\baselineskip}\noindent\ignorespacesafterend}

\usepackage{enumitem}

\acmConference{}{}{}
\title{Game Hopping in Lean}
\author{Stefan Dziembowski}
\affiliation{%
  \institution{University of Warsaw}
  \city{Warsaw}
  \country{Poland}
}
\affiliation{%
  \institution{IDEAS Research Institute}
  \city{Warsaw}
  \country{Poland}
}

\author{Grzegorz Fabia\'nski}
\affiliation{%
  \institution{IDEAS Research Institute}
  \city{Warsaw}
  \country{Poland}
}

\author{Daniele Micciancio}
\affiliation{%
  \institution{University of California San Diego}
  \city{La Jolla}
  \country{USA}
}

\author{Rafa\l{} Stefa\'nski}
\affiliation{%
  \institution{IDEAS Research Institute}
  \city{Warsaw}
  \country{Poland}
}

\begin{document}

\begin{abstract}
We present HOPSCOTCH, a Lean 4 framework for mechanizing computationally sound, game-based cryptographic proofs. Security definitions are expressed as indistinguishability between stateful probabilistic oracles, and proofs follow the standard game-hopping paradigm. HOPSCOTCH uses a shallow embedding: oracles and reductions are ordinary Lean definitions, enabling direct integration with the full Lean ecosystem, including general mathematical theories from Mathlib, such as finite-group theory. A game-hopping proof in HOPSCOTCH is represented as an explicit formal object whose constructors correspond to the standard steps of a game-hopping argument, making proofs easier to construct, automate, and inspect. We prove a general computational soundness theorem that interprets these proof objects by constructing reductions against the assumptions they use and deriving a concrete bound on the advantage of any distinguisher. Observational equivalence between oracles is established using a state-abstraction methodology: a simple yet powerful approach that supports transformations such as adding or forgetting state and replacing eager sampling with lazy sampling. We illustrate the framework with formalized proofs of the IND-CCA security of encrypt-then-MAC, the security of ElGamal encryption from DDH, the implication from one-time secrecy to public-key IND-CPA security, and the GGM pseudorandom-function construction. To the best of our knowledge, the last is the first mechanized proof of GGM for non-constant depth.
\end{abstract}

\maketitle

\section{Introduction}
\label{sec:introduction}

Modern cryptography relies on the paradigm of provable security, typically formalized via computational game-based definitions. In this framework, the security of a primitive or protocol is expressed as a game played between a challenger and a computationally bounded adversary. Proving security involves constructing a sequence of game-based reductions -- often referred to as game hopping or hybrid arguments -- that bound the adversary's advantage by reducing it to one or more known hard computational problems or basic cryptographic primitives.

While conceptually elegant, manual game-based proofs are notoriously error-prone, monolithically complex, and frequently plagued by subtle gaps. This has motivated a rich line of research into formal methods for cryptographic security analysis that offer the strong security guarantees of computational security definitions together with the reliability of mechanically verified and partially automated proofs. Domain-specific tools such as EasyCrypt \cite{EasyCrypt} or CryptoVerif \cite{blanchet2008computationally} have successfully automated large classes of game-based reductions. However, these specialized verification tools operate outside of general-purpose interactive theorem provers, forcing a stark trade-off: developers must choose between the automated convenience of a domain-specific tool and the foundational expressiveness and foundational soundness of a general-purpose proof assistant.

\subsection{Our contribution}
In this paper, we bridge this gap by introducing HOPSCOTCH\footnote{This name stands for \emph{Hybrid Oracle Proofs of Security via a Calculus of
Optimized Tactics for Cryptographic Hopping}. It also refers to
the children's game: cryptographic game-hopping proofs likewise proceed
through a sequence of carefully arranged hops, coin tosses,
and, at least occasionally, some fun.
Notably missing from the children's game is the presence of an adversary.},
a computationally sound Lean 4 \cite{Moura2021} framework for mechanizing game-based
cryptographic proofs. The project’s source code is available under the MIT License on GitHub\footnote{
    \url{https://github.com/ravst/Hopscotch}
}.
HOPSCOTCH is designed to follow the structure of
pen-and-paper security proofs written in the game-hopping paradigm
\cite{shoup2004sequences, rosulek2021joy}: the user defines a sequence of
games (oracles), justifies each hop either by an observational-equivalence argument
or by a reduction to a cryptographic assumption, and obtains a
mechanically checked reductionist security statement.
The distinguishing features of HOPSCOTCH are the following.

\begin{enumerate}
\item \textbf{Security definitions as oracle indistinguishability.}
Cryptographic security notions are expressed directly as
indistinguishability statements between stateful probabilistic oracles.
This matches the usual game-based presentation of cryptography: an
adversary interacts with one of two oracles, and security means that no
efficient adversary can tell which oracle it is interacting with.


\item \textbf{Proofs as explicit formal objects.}
A HOPSCOTCH proof is not merely an unstructured Lean script proving an
indistinguishability theorem.  It is an explicit proof object whose
constructors correspond to the steps of a cryptographic proof:
reflexivity, symmetry, transitivity, observational equivalence,
reduction hops, and hybrid sequences. The structure of such a
proof object mirrors the structure of a game-hopping argument.

\item \textbf{A computational interpretation of proof objects.}
We prove a general computational soundness theorem for these proof
objects.  A HOPSCOTCH proof records the structure of a game-hopping
argument independently of any particular adversary.  The soundness theorem
then gives this adversary-free proof object its computational meaning:
for every distinguisher against the endpoint oracles, it constructs the
corresponding distinguishers against the assumptions used in the proof,
and bounds the original advantage in terms of their advantages.

\item \textbf{A shallow embedding in Lean.}
HOPSCOTCH does not introduce a separate programming language for games.
Oracles, reductions, and adversaries are ordinary Lean definitions,
written using Lean's existing syntax and type system.  This allows users
to rely directly on the surrounding Lean ecosystem, including standard
mathematical libraries, tactic automation, and reasoning
principles, rather than translating cryptographic games into a restricted
domain-specific language.  Moreover, domain-specific game-hopping
automation can be freely interleaved with ordinary Lean proof
development: every game-hopping proof obligation can be discharged
using the full power of Lean.  In this sense,
HOPSCOTCH combines the convenience of a specialized game-hopping
framework with the expressiveness of a general-purpose proof assistant.

\item \textbf{Support for parameter-dependent hybrid arguments.}
The framework supports hybrid sequences whose length is not fixed in
advance, but may depend on the security parameter or on an explicit
bound on the number of oracle queries made by the adversary.  This is
essential for formalizing many standard cryptographic arguments.

\item \textbf{Observational equivalence by abstraction.}
For hops that are purely semantic, HOPSCOTCH provides an abstraction
methodology for proving observational equivalence between oracles.
Instead of requiring users to reason in a relational program logic, the
user supplies an abstraction function explaining how the state of one
oracle represents the state of another.  This logic-free technique
captures common proof patterns such as forgetting irrelevant state,
introducing redundant state, reshuffling state, and replacing eager
sampling with lazy sampling.

\item \textbf{Domain-specific proof automation.}
HOPSCOTCH provides tactics that automate recurring steps in game-hopping
proofs, reducing the amount of boilerplate and making formal proofs shorter
and easier to read. This automation is
enabled in part by the explicit proof-object representation, which gives the
tactics a structured target to build rather than requiring them to manipulate
an arbitrary Lean proof state.

\end{enumerate}

Together, these features allow HOPSCOTCH to cover a range of proof
patterns that arise in standard cryptography: fixed sequences of games,
reductions to cryptographic assumptions, query-bounded hybrids, lazy
random-function implementations, and hybrid arguments whose length grows
with the security parameter.  We demonstrate this by mechanizing several
representative proofs, including the IND-CCA security of the
encrypt-then-MAC construction (Section~\ref{subsec:encrypt-then-mac}), the security of ElGamal encryption from
DDH (Section~\ref{subsec:elgamal}), the implication from one-time secrecy to public-key IND-CPA
security (Section~\ref{subsec:ots-implies-ind-cpa}), and the GGM construction of pseudorandom functions (Section~\ref{subsec:ggm}).
To the best of our knowledge, this is the first mechanized proof
of the GGM construction of non-constant depth (with \cite{brzuska2024cryptozoo} formalizing
the GGM construction of depth $3$).

\subsection{Related work}

As mentioned above, we are mainly interested in frameworks grounded in
general-purpose theorem provers.  Nevertheless, we would like to mention
ProofFrog~\cite{evans2024prooffrog}, a domain-specific tool that was one
of the inspirations for HOPSCOTCH. ProofFrog presents
proofs as explicit sequences of oracle games, with each hop justified by
an assumption or by equivalence of adjacent games.  The main difference
is foundational: ProofFrog uses a dedicated C/Java-style language and a
custom automatic checker based on syntactic equivalence heuristics,
whereas HOPSCOTCH represents oracles as ordinary Lean definitions and
justifies transformations by Lean theorems checked by Lean's kernel.
This makes ProofFrog lightweight and more automatic, while HOPSCOTCH
offers a fallback to the full Lean environment when automation is not
enough.

The work most closely related to ours is VCVio~\cite{VCVio,vcvioOld}, a framework for formalizing cryptographic proofs in Lean. One subtle point is that the two cited versions of VCVio address somewhat different goals. The earlier version~\cite{vcvioOld} provides a framework for modeling abstract oracle computations using free monads, whereas the more recent version~\cite{VCVio} supports reasoning about the indistinguishability of stateful oracles. We initially developed HOPSCOTCH on top of the earlier version, while the newer version was developed concurrently with our work. After its release, we ported HOPSCOTCH to the new version, both to maintain compatibility and to benefit from its improved support for abstract oracle computations.
Other foundational cryptographic frameworks in general-purpose theorem provers include CertiCrypt~\cite{CertiCrypt}, FCF~\cite{fcf}, and SSProve~\cite{SSProve} in Rocq, and CryptHOL~\cite{CryptHOL} in Isabelle/HOL. These systems provide rich languages and logics for constructing kernel-checked cryptographic proofs, using techniques such as probabilistic relational Hoare logic. Compared with these frameworks, including the recent version of VCVio, HOPSCOTCH provides two complementary features.


First, it represents the
domain-specific structure of a game-hopping argument as an explicit Lean
object.  Its constructors record equivalence hops, uses of assumptions,
reductions, transitivity, and parameterized hybrid sequences.  A generic
soundness theorem gives this object a computational interpretation.  For
each cryptographic assumption appearing in the proof, it constructs a
single adversary against that assumption, regardless of how many times the
assumption is used --- the number of uses appears only as a multiplicative
factor in the resulting concrete security bound.  This is essential for our
semi-formal\footnote{i.e., based on external polynomial-time assumptions verified manually by the reader; see Section~\ref{subsec:asymptotic-computational-soundness} for details.}
approach to establishing correct asymptotics for hybrid arguments whose length depends on
the security parameter. In particular, the GGM security proof and the proof of
public-key IND-CPA security from one-time secrecy fall into this category. To
the best of our knowledge, we present the first framework with this functionality.
Moreover, as mentioned before, the definition of an explicit proof object facilitates
domain-specific proof automation.

Second, HOPSCOTCH provides an abstraction-based method for proving
observational equivalence of oracle implementations.  The user supplies
a function mapping the state of one oracle to the state of another and
proves that it relates the initial states and commutes with every query.
This approach often produces short proofs for changes of representation,
duplicated or forgotten state, and delayed randomness.
In particular, the randomized version of the abstraction allows
relating implementations that sample randomness eagerly and lazily.
This is a problem that is difficult to solve in pRHL, as attested
by \cite{metere2023direct}, where the authors present a solution
by extending the syntax of EasyCrypt.

Finally, we would like to mention that, similarly to most other tools, HOPSCOTCH does not verify the polynomial-time complexity of the reductions used in the proof.
Instead, we rely on the concrete security bounds, or on verifying this external condition by the user.
See~\ref{subsec:computational-soundness} for more details.

\paragraph{Use of AI tools.}
We used AI-assisted tools during the development of the artifact and the
preparation of the manuscript. For the Lean code, we used Codex with
GPT-5.5~\cite{singh2025openai}, DeepSeek-V4 \cite{deepseekV4} and Aristotle by
Harmonic~\cite{achim2025aristotleimolevelautomatedtheorem}. Codex was particularly helpful in
developing tactics and proving small auxiliary lemmas. Aristotle was used
in a more autonomous mode: given the statement that every observationally
equivalent pair of oracles factors through two abstractions, it constructed
a proof together with several intermediate definitions and lemmas. For the
manuscript, we used GPT-5.5 and Grammarly for rephrasing, editing,
and spelling and grammar correction.

\section{Framework}
\label{sec:framework}
In this section, we give an overview of our game-hopping framework in Lean.
Section~\ref{subsec:oracle-indistinguishability-example} introduces our
motivating example: the proof that IND-RAND security of an encryption scheme
implies its IND-CPA security. Section~\ref{subsec:basic-components} describes
the basic components of our Lean framework, including cryptographic schemes,
oracles, and oracle reductions.
Section~\ref{subsec:proof-principles-and-automation} explains how formal oracle
indistinguishability proofs can be constructed in our framework, how this
construction is automated by domain-specific tactics, and how the
example IND-RAND-to-IND-CPA proof is formalized. Finally,
Section~\ref{subsec:computational-soundness} discusses the computational
interpretation of our formal indistinguishability proofs.

\subsection{Oracle Indistinguishability by Example}
\label{subsec:oracle-indistinguishability-example}
Many cryptographic security definitions can be understood as games between an
adversary and a hidden system. The adversary may interact with the system by
making queries and observing the replies, but should still be unable to learn
which of two possible systems it is talking to. In this section, we model these
systems as \emph{oracles}: stateful randomized procedures that answer queries
and may update their internal state. Security is then expressed as
indistinguishability of oracles.

For example, consider an encryption scheme consisting of the following
components:
\[
\Pi_\kappa =
\left(
\begin{array}{rcl}
\mathsf{KeyGen} & : & \mathsf{Dist}(\mathcal{K}_\kappa) \\
\mathsf{Enc} & : & \mathcal{K}_\kappa \times \mathcal{M}
  \to \mathsf{Dist}(\mathcal{C}) \\
\mathsf{Dec} & : & \mathcal{K}_\kappa \times \mathcal{C}
  \to \mathcal{M}
\end{array}
\right).
\]
Here, $\mathsf{Dist}$ denotes a probability distribution, $\mathcal{M}$ denotes the space of plaintext messages, $\mathcal{C}$
denotes the space of ciphertexts, and $\mathcal{K}_\kappa$ denotes the set of
keys for security parameter\footnote{
  The security parameter is usually denoted as $\lambda$, but we use $\kappa$ as $\lambda$ is a reserved keyword in Lean.
} $\kappa \in \Nat$. It is common to treat a
cryptographic scheme as a family of algorithms indexed by such a security
parameter. 
In the asymptotic approach to computational security, this parameter is used to
measure the adversary's running time and its ability to distinguish the two
systems: typically, the adversary's running time should be polynomial in
$\kappa$, while its probability of distinguishing the two oracles correctly
should exceed random guessing only by a negligible function of $\kappa$.
In the case of encryption schemes, the security parameter
usually controls the length of the key. In practice, the scheme is
instantiated with a single sufficiently large value of $\kappa$.

A standard security notion for encryption schemes is indistinguishability under
chosen-plaintext attack, or IND-CPA. It can be expressed as indistinguishability
of two oracles, which we call $\mathsf{IndCpaLeft}_\kappa$ and
$\mathsf{IndCpaRight}_\kappa$. Both oracles have an internal state consisting of
a secret key $k \in \mathcal{K}_\kappa$, sampled initially according to
$\mathsf{KeyGen}$. The two oracles accept the same kind of query: a pair of
messages $(m_0,m_1)$ of equal length. The left oracle responds by encrypting the
first message, sampling the value from $\mathsf{Enc}(k,m_0)$. The right oracle responds by
encrypting the second message, sampling the value from $\mathsf{Enc}(k,m_1)$. The security
claim is that $\mathsf{IndCpaLeft}_\kappa$ and
$\mathsf{IndCpaRight}_\kappa$ are hard to distinguish, even though the adversary
chooses the two candidate plaintexts in each query.

Another common security notion for encryption schemes is real-or-random
security, or IND-RAND. This notion compares a real encryption oracle with a
random ciphertext oracle. Unlike the IND-CPA oracles above, these oracles accept
queries consisting of a single message. The real oracle,
$\mathsf{EncReal}_\kappa$, stores a secret key
$k \in \mathcal{K}_\kappa$ sampled according to $\mathsf{KeyGen}$ and, on query
$m$, returns a ciphertext sampled from $\mathsf{Enc}(k,m)$. The oracle
$\mathsf{EncRand}_\kappa$, accepts the same query $m$, but ignores the contents
of the message and returns a uniformly random ciphertext\footnote{This is a common, but slightly stronger security definition compared to the standard IND-CPA. Our example is easily adapted to a definition perfectly equivalent to IND-CPA by replacing the random ciphertext with an encryption of a fixed message $\mathsf{Enc}(k,0^{|m|})$ of the same length as $m$.} of the appropriate
length. The IND-RAND security claim is that these two oracles are hard to
distinguish: encryptions of chosen messages should look like uniformly random
ciphertexts. 

To illustrate the game-hopping proof strategy, let us present an informal
overview of the proof that IND-RAND security implies IND-CPA security.
For this, we require one more object, called
an \emph{oracle reduction}, which is one of the main ingredients of a
game-hopping proof. Intuitively, an oracle reduction is itself an oracle, but
instead of computing its replies on its own, it may query another oracle.
We write the composition of an oracle reduction $R$ with an oracle
$\mathcal{O}$ using the diamond notation $R \diamond \mathcal{O}$. This denotes
the oracle obtained by running $R$ while giving it access to $\mathcal{O}$ as its
underlying oracle.

We are now ready to present the game-hopping proof. The proof uses two simple
oracle reductions, $R_L$ and $R_R$. Both reductions expose the IND-CPA query
interface: they accept a pair $(m_0,m_1)$ of equal-length messages and return a
ciphertext.
They use access to an underlying oracle with the IND-RAND query interface, which
takes single messages as queries and returns ciphertexts.
The reduction $R_L$ answers the query by forwarding $m_0$ to this
underlying oracle and returning the ciphertext it receives. The reduction $R_R$
does the same with $m_1$. If we combine $R_L$ with the real encryption oracle
$\mathsf{EncReal}_\kappa$, the resulting oracle behaves exactly like
$\mathsf{IndCpaLeft}_\kappa$: it samples a key once, and on each query
$(m_0,m_1)$, it returns an encryption of $m_0$. Since two oracles with the same
behavior are clearly indistinguishable, we obtain the following indistinguishability relation
(where, \(\approx\) denotes the intuitive notion of oracle
indistinguishability -- this notation is used only in this informal example):
\[
\mathsf{IndCpaLeft}_\kappa
\approx
R_L \diamond \mathsf{EncReal}_\kappa.
\]
By the IND-RAND assumption, $\mathsf{EncReal}_\kappa$ is indistinguishable from
$\mathsf{EncRand}_\kappa$. Indistinguishability is preserved under oracle
reductions: if an adversary could distinguish
$R_L \diamond \mathsf{EncReal}_\kappa$ from
$R_L \diamond \mathsf{EncRand}_\kappa$, then we could build an adversary
distinguishing $\mathsf{EncReal}_\kappa$ from $\mathsf{EncRand}_\kappa$ by
simulating the reduction $R_L$. Hence:
\[
R_L \diamond \mathsf{EncReal}_\kappa
\approx
R_L \diamond \mathsf{EncRand}_\kappa.
\]
Now consider $R_L \diamond \mathsf{EncRand}_\kappa$. On input $(m_0,m_1)$, it
queries the random ciphertext oracle on $m_0$. But
$\mathsf{EncRand}_\kappa$ ignores the contents of its input and returns a
uniformly random ciphertext of the appropriate length. Since IND-CPA queries
require $m_0$ and $m_1$ to have the same length, querying
$\mathsf{EncRand}_\kappa$ on $m_0$ has the same observable behavior as querying
it on $m_1$. Thus:
\[
R_L \diamond \mathsf{EncRand}_\kappa
\approx
R_R \diamond \mathsf{EncRand}_\kappa.
\]
Applying the IND-RAND assumption once more, now through the reduction $R_R$,
gives:
\[
R_R \diamond \mathsf{EncRand}_\kappa
\approx
R_R \diamond \mathsf{EncReal}_\kappa.
\]
Finally, combining $R_R$ with the real encryption oracle gives an oracle that
behaves exactly like $\mathsf{IndCpaRight}_\kappa$: it samples a key once, and
on each query $(m_0,m_1)$, it returns an encryption of $m_1$. Therefore:
\[
R_R \diamond \mathsf{EncReal}_\kappa
\approx
\mathsf{IndCpaRight}_\kappa.
\]
Putting these steps together yields the following chain of game hops:
\[
\mathsf{IndCpaLeft}_\kappa
\approx
R_L \diamond \mathsf{EncReal}_\kappa
\approx
R_L \diamond \mathsf{EncRand}_\kappa
\approx
R_R \diamond \mathsf{EncRand}_\kappa
\approx
R_R \diamond \mathsf{EncReal}_\kappa
\approx
\mathsf{IndCpaRight}_\kappa.
\]
By transitivity of indistinguishability, $\mathsf{IndCpaLeft}_\kappa$ and
$\mathsf{IndCpaRight}_\kappa$ are indistinguishable. Thus, assuming the real
encryption oracle is indistinguishable from the random ciphertext oracle, the
encryption scheme is IND-CPA secure.

\subsection{Basic Components}
\label{subsec:basic-components}

We now describe how the game-hopping framework is represented in Lean,
starting with its basic components: schemes, oracles, and oracle reductions.

\subsubsection{Schemes}

A cryptographic scheme may be any Lean object from which the relevant oracles will
be constructed. However, it is usually most convenient to package
the algorithms of a scheme as fields of a record. For instance, we represent a
symmetric encryption scheme as presented\footnote{The Lean listings are lightly simplified for readability, primarily by omitting implicit arguments, namespace qualifiers, typeclass instances, and inessential type annotations. As a result, the displayed snippets are illustrative and may not parse as written.
 We hope that their meaning is intuitively clear even to readers unfamiliar with Lean.
 The complete code is available in the project repository.}
in Figure~\ref{fig:sym-enc-scheme}.

\begin{figure}[H]
\centering
\begin{minipage}{\linewidth}
\begin{lstlisting}[
  language=lean,
  aboveskip=0.75\baselineskip
]
structure SymEncScheme (K : Type) (C : (msg_len : ℕ) → Type) where
  keyGen : PMF K
  encrypt : K → BitVec n → PMF (C n)
  decrypt : K → C n → BitVec n

structure SymEncSchemeFam (K : (κ : ℕ) → Type) (C : (κ : ℕ) → (n : ℕ) → Type) where
  scheme : (κ : ℕ) → SymEncScheme (K κ) (C κ)
\end{lstlisting}
\end{minipage}
\caption{Lean representation of symmetric encryption schemes and scheme families.}
\label{fig:sym-enc-scheme}
\end{figure}

In this model, keys are elements of an abstract parameter type
\lstinline{K}, messages are bit vectors \lstinline{BitVec n} indexed by their
length \lstinline{n}, and ciphertexts are elements of a family of types
\lstinline{C n}, also indexed by the plaintext length. Thus, the encryption of
an \lstinline{n}-bit message has type \lstinline{C n}. (This length-indexed
representation will later allow the oracle $\mathsf{EncRand}$ to sample a
ciphertext matching the length of the given plaintext.) The field
\lstinline{keyGen} is a probability distribution from which keys should be sampled,
and \lstinline{encrypt} is a randomized encryption algorithm. We also include
the decryption algorithm for completeness, although it does not play a role in
the security definitions. Randomness is represented using the
probability mass function type \lstinline{PMF}. Finally, \lstinline{SymEncSchemeFam}
lifts this representation to families of schemes indexed by the security parameter $\kappa$.

\subsubsection{Oracles}

When modeling oracles, we separate their \emph{specifications} from their
\emph{implementations}. The specification fixes the oracle interface: it
describes which queries may be asked and what type of answer each query returns.
The implementation describes how the oracle answers these queries. For example,
the IND-CPA oracles $\mathsf{IndCpaLeft}_\kappa$ and
$\mathsf{IndCpaRight}_\kappa$ have the same specification, but different
implementations.

To represent oracle specifications, we use the \lstinline{OracleSpec} type from
the VCVio library. It is parameterized by a type \lstinline{ι} of queries, and
assigns a return type to each query:
\begin{leandisplay}
\leanline!def OracleSpec (ι : Type) := ι → Type!
\end{leandisplay}
Thus, for a specification \lstinline{S : OracleSpec ι}, an element
\lstinline{i : ι} represents an oracle query, and \lstinline{S i} is the type of
the corresponding answer. For example, the IND-CPA oracle interface can be
represented by a query type with one constructor, containing a length
\lstinline{n} and two messages of this length. The answer type for such a query
is \lstinline{C n}, as shown in Figure~\ref{fig:ind-cpa-spec}.
\begin{figure}[H]
\centering
\begin{minipage}{\linewidth}
\begin{lstlisting}[
  language=lean,
  aboveskip=0.75\baselineskip
]
inductive IndCpaDomain : Type
| eavesdrop (n : ℕ) (msgs : (BitVec n) × (BitVec n))

def IndCpaSpec (C : ℕ → Type) : OracleSpec IndCpaDomain :=
  fun ⟨n, _⟩ => C n
\end{lstlisting}
\end{minipage}
\caption{Lean representation of the IND-CPA oracle specification.}
\label{fig:ind-cpa-spec}
\end{figure}

An oracle implementation consists of an internal state type, an initial-state
distribution, and a query implementation. Given a query and the current state,
the query implementation returns a distribution over pairs consisting of an
answer and an updated state. We represent such stateful probabilistic
computations using the monadic type
\begin{leandisplay}
\leanline!abbrev RState (σ : Type) (α : Type) : Type := StateT σ PMF α!
\end{leandisplay}
The type \lstinline{StateT σ M α} is defined as
\lstinline{σ → M (α × σ)}, so it directly captures a computation that transforms
a state while producing a value. In our case, we instantiate \lstinline{M} with
the \lstinline{PMF} monad to account for probabilistic choice. Thus, for an
oracle specification \lstinline{S : OracleSpec ι} and state type
\lstinline{σ}, a query implementation has type
\begin{leandisplay}
\leanline!(q : ι) → RState σ (S q)!
\end{leandisplay}
VCVio packages such functions as \lstinline{QueryImpl S (RState σ)}, which we
use in the definition of a random stateful oracle shown in
Figure~\ref{fig:rstate-oracle}.
\begin{figure}[H]
\centering
\begin{minipage}{\linewidth}
\begin{lstlisting}[
  language=lean,
  aboveskip=0.75\baselineskip
]
structure OracleImpl {I : Type u} (O : OracleSpec I) where
  stateType : Type u
  initialState : PMF stateType
  queries : QueryImpl O (RState stateType)
\end{lstlisting}
\end{minipage}
\caption{Lean representation of the random stateful oracle.}
\label{fig:rstate-oracle}
\end{figure}

Thanks to the monadic representation, query implementations can be written in a
style similar to imperative programming languages, using \lstinline{do}
notation. For example, see Figure~\ref{fig:ind-cpa-left}
for the implementation of the $\mathsf{IndCpaLeft}$ oracle.
The $\mathsf{EncReal}$ and $\mathsf{EncRand}$ oracles are
implemented similarly; see
\path{GameHoppingInLean/Examples/SecurityDefinitions/IndCpaRand.lean} in the
accompanying Lean code.

\begin{figure}[H]
\centering
\begin{minipage}{\linewidth}
\begin{lstlisting}[
  language=lean,
  aboveskip=0.75\baselineskip
]
def IndCpaL (scheme : SymEncScheme K C) : OracleImpl (IndCpaSpec C) where
  stateType := K
  initialState := scheme.keyGen
  queries := fun ⟨_, (m₀, _)⟩ => do
    /- Retrieve the secret key from the oracle state. -/
    let key <- get
    /- Sample the ciphertext from the possible encryptions of the left message. -/
    let c <- scheme.encrypt key m₀
    /- Return the ciphertext as the oracle output. -/ 
    return c

-- Family of oracles, parameterized by the security parameter κ.
def IndCpaLFam (schemeFam : SymEncSchemeFamily K C) :=
  fun κ => IndCpaL (schemeFam.scheme κ)

  
\end{lstlisting}
\end{minipage}
\caption{Lean implementation of the IND-CPA left oracle.}
\label{fig:ind-cpa-left}
\end{figure}


\subsubsection{Oracle reductions}

The next component of a game-hopping proof are \emph{oracle reductions}. An
oracle reduction is similar to an oracle: it consists of an internal state, an
initialization procedure, and query implementations. Unlike a simple oracle,
however, it might query another oracle of a fixed specification. Since
only the specification of this underlying oracle is known, the reduction must
be prepared to work with any implementation of the specification.
To represent such computations, we use the VCVio type
\lstinline{OracleComp S α}. This is an abstract computation that may query an
oracle of specification \lstinline{S} and returns a value of type
\lstinline{α}. VCVio implements these computations using free monads. In
addition to querying the underlying oracle, a reduction must also be able to
access its own state and perform probabilistic sampling. We model these effects
as abstract operations as well: \lstinline{get} and \lstinline{put} for state,
and \lstinline{sample} for probabilistic choice. For this purpose, we define
\lstinline{withPMFAndState}, which takes a specification \lstinline{S} and
extends it with the state and sampling operations; see
Figure~\ref{fig:with-pmf-and-state}.
\begin{figure}[H]
\centering
\begin{minipage}{\linewidth}
\begin{lstlisting}[
  language=lean,
  aboveskip=0.75\baselineskip
]
/- The list of allowed operations for a reduction -/
inductive withPMFAndStateI (I : Type) (S : Type) : Type
  | oracle (i : I) -- Send the query `i` to the underlying oracle
  | sample (α : Type) (d : PMF α) -- Sample a value from the distribution `d`
  | getState -- Get the current state of the reduction
  | setState (st : S) -- Update the state of the reduction to `st`

/- To define the specification, we specify the output types for each operation -/
def withPMFAndStateSpec (S : Type) (O : OracleSpec I) :
    OracleSpec (withPMFAndStateI I S)
  | .oracle i => O i
  | .withPMFAndStateI.sample α _ => α
  | .getState => S
  | .setState _ => Unit
\end{lstlisting}
\end{minipage}
\caption{Extending the specification with state and sampling operations.}
\label{fig:with-pmf-and-state}
\end{figure}
Initialization uses a slightly smaller effect signature. Since the reduction
state has not yet been created, the initializer may query the underlying oracle
and sample random values, but it cannot access or update the reduction state.
We capture this using \lstinline{withPMF}, a variant of
\lstinline{withPMFAndState} that extends \lstinline{S} only with the sampling
operation. With these definitions, an oracle reduction has essentially the same structure as an oracle
implementation, except that its initialization and query procedures are written
in the appropriate \lstinline{OracleComp} monads rather than in \lstinline{RState}; see Figure~\ref{fig:oracle-reduction}.
\begin{figure}[H]
\centering
\begin{minipage}{\linewidth}
\begin{lstlisting}[
  language=lean,
  aboveskip=0.75\baselineskip
]
/-- An oracle reduction is parametrized by both its input and output specifications. -/ 
structure OracleReduction (O₁ : OracleSpec I₁) (O₂ : OracleSpec I₂) where
  stateType : Type
  initialState : OracleComp (withPMFSpec O₁) stateType
  queries : QueryImpl O₂ (OracleComp (withPMFAndStateSpec stateType O₁))
\end{lstlisting}
\end{minipage}
\caption{Definition of an oracle reduction.}
\label{fig:oracle-reduction}
\end{figure}

For example, Figure~\ref{fig:ind-rand-reductions} shows the oracle reduction
$R_L$ used in our running IND-RAND-to-IND-CPA example ($R_R$ is
defined analogously). This first reduction is deliberately simple: in particular,
it has no internal state. Later in the paper, we will consider more involved examples.
\begin{figure}[H]
\centering
\begin{minipage}{\linewidth}
\begin{lstlisting}[
  language=lean,
  aboveskip=0.75\baselineskip
]
def IndCpaRand_to_IndCpaL : OracleReduction (IndCpaRandSpec C) (IndCpaSpec C) where
  stateType := Unit
  initialState := return ()
  queries := fun ⟨n, (m₀, _m₁)⟩ =>
    -- Pass m₀ to the underlying oracle, and forward its output
    OracleReduction.query (IndCpaRandDomain.ctxt n m₀)
\end{lstlisting}
\end{minipage}
\caption{Definition of the RL reduction.}
\label{fig:ind-rand-reductions}
\end{figure}

Finally, the key property of oracle reductions is that
they can be composed with concrete oracle implementations, yielding another concrete
oracle implementation. The state of the composed oracle is the product of
the reduction state and the state of the underlying oracle.
Both the initialization and query procedures of the composed oracle are obtained
using VCVio's \lstinline{simulateQ} operation. This operation interprets an
abstract computation of type \lstinline{OracleComp S α} using a concrete query
implementation of type \lstinline{QueryImpl S M}, producing a computation of
type \lstinline{M α}. In this case, we instantiate \lstinline{M} with
\lstinline{RState (σ × τ)}, the monad of stateful probabilistic computations
that keeps track of both the reduction state, of type \lstinline{σ}, and the
underlying oracle state, of type \lstinline{τ}.
We implement this operation as the function:
\begin{leandisplay}
\leanline!apply : OracleReduction O₁ O₂ → OracleImpl O₁ → OracleImpl O₂!
\end{leandisplay}
and introduce the diamond notation \lstinline{R ◇ O} for it. The definition can
be found in \path{GameHoppingInLean/OracleReductions.lean} in the accompanying Lean code.


\subsection{Indistinguishability Proofs}
\label{subsec:proof-principles-and-automation}

We now describe the reasoning rules used to construct formal
indistinguishability proofs. These rules correspond to the main steps of a
game-hopping argument: replacing an oracle by an observationally equivalent one,
invoking a cryptographic assumption, applying an oracle reduction, and composing
consecutive hops. We then describe the domain-specific tactics that automate
parts of the construction of the indistinguishability proofs,
and explain how we formalize the IND-RAND to IND-CPA argument from
Section~\ref{subsec:oracle-indistinguishability-example}.

\subsubsection{Observational equivalence}
In this section, we formalize the idea that two oracle implementations
\emph{behave in the same way}---a notion used to justify some of the hops in
our example proof. It captures the situation where two oracles may differ in
their internal implementation, but expose the same observable behavior. In this
paper, we refer to this property as \emph{observational equivalence}.
Intuitively, two oracles are observationally equivalent if no external observer,
even a computationally unbounded one, can distinguish them by interacting with them.

We define observational equivalence using query sequences. First, we introduce
the procedure \lstinline{runQueries}, which inputs a finite list of queries to
an oracle (one by one) and returns the resulting distribution over lists of outputs.
We then say that two oracles are observationally equivalent (\lstinline{ObsEq})
if, for every finite list of queries, running the two oracles on that list produces the same
distribution over output lists. (The Lean implementation of \lstinline{runQueries} and \lstinline{ObsEq} are
straightforward, so we do not include them as separate code snippets here; the
definitions can be found in the accompanying Lean code).

At first sight, this query-sequence formulation may seem weaker than an
interactive one, since the entire list of queries is fixed in advance. In our
setting, however, this causes no loss of generality. Observational equivalence
requires exact equality of output distributions, rather than an approximate
notion of distributional equivalence, such as computational indistinguishability
or statistical closeness. As a result, the query-sequence and interactive
formulations coincide: any adaptive interaction can be decomposed according to
its possible query-output histories, and equality for all fixed query sequences
implies equality of the observer's final output distribution. We have formalized
this interactive--non-interactive equivalence in Lean as part of the
computational soundness theorem, see lemma \lstinline{obsEq_eq_advantage_zero}. 

Proving observational equivalence directly from the definition can be
cumbersome, because it requires reasoning about arbitrary finite query
sequences.  For this reason, our framework provides two local proof principles,
both based on the idea of abstracting oracle states called
\emph{deterministic state abstraction} and \emph{randomized state abstraction}.
Let us start by explaining the simpler, deterministic version.
Let \lstinline{O1} and \lstinline{O2} be oracles with the same specification, but
possibly different state types \lstinline{S1} and \lstinline{S2}.  A
deterministic state abstraction from \lstinline{O1} to \lstinline{O2} is a
function \lstinline{f : S1 -> S2} satisfying two conditions.  First, applying
\lstinline{f} to the initial-state distribution of \lstinline{O1} gives the
initial-state distribution of \lstinline{O2}.  Second, the abstraction is
preserved by every query: for every query \lstinline{q} and state
\lstinline{s1 : S1}, running \lstinline{O1} from \lstinline{s1} and then
applying \lstinline{f} to the updated state gives the same distribution over
replies and abstract states as first applying \lstinline{f} to
\lstinline{s1} and then running \lstinline{O2}.  The formal statement is shown
in Figure~\ref{fig:state-abstraction}.

\begin{figure}[H]
\centering
\begin{minipage}{\linewidth}
\begin{lstlisting}[
  language=lean,
  aboveskip=0.75\baselineskip
]
def mapSecond (f : β → γ) (p : α × β) : α × γ := (p.1, f p.2)

def mapInputState (f : S₁ → S₂) (m : RState S₂ α) (s : S₁) : PMF (α × S₂) :=
  StateT.run m (f s)

def mapOutputState (f : S₁ → S₂) (m : RState S₁ α) (s : S₁) : PMF (α × S₂) :=
  (StateT.run m s).map (mapSecond f)

def correctAbstraction (ro₁ ro₂ : OracleImpl O) (f : ro₁.stateType → ro₂.stateType) :=
  ro₁.initialState.map f = ro₂.initialState ∧
  ∀ (query : O.Domain),
      mapOutputState f (ro₁.queries query) =
      mapInputState f (ro₂.queries query)
\end{lstlisting}
\end{minipage}
\caption{Deterministic state abstraction.}
\label{fig:state-abstraction}
\end{figure}

The key property of a correct abstraction is that it implies observational
equivalence, as formalized by the theorem \lstinline{correctAbstractionImpliesObsEq}.
It can be shown by induction on
the query sequence, while maintaining the joint distribution of the outputs
produced so far and the current oracle state; treating these distributions
separately would lose the necessary correlation.

Let us now discuss the \emph{randomized state abstraction}, in which the
state map has type \lstinline{S1 -> PMF S2}.  Its correctness conditions have
the same shape as in the deterministic case, except that \lstinline{PMF.map},
which applies a deterministic function to a distribution, is replaced by
\lstinline{PMF.bind}, which works with probabilistic functions.
As in the deterministic case, the existence of a correct randomized abstraction
implies observational equivalence; this is formalized as
\lstinline{correctAbstractionBindImpliesObsEq}.
It might be worth noting that the randomized version generalizes the deterministic one,
which can be recovered by lifting 
 \lstinline{f : S1 -> S2} to
\lstinline{fun s1 => pure (f s1)}. However, it is convenient to keep
both principles in the framework. 
The deterministic version is simpler
to use when it applies, while the randomized version is more expressive.

The state abstraction technique is surprisingly expressive and easy to work
with. In fact, it is the only proof principle we use in our examples to prove
observational equivalence. Such equivalence proofs often require relating two
program executions, a task commonly supported by relational program logics such
as pRHL. For the examples considered in this paper, however, the abstraction
principle itself was sufficient to capture the required relationships between
the two oracle states.
As we will see in Section~\ref{sec:case-studies}, deterministic abstractions can
express state invariants, such as the fact that two fields of the
\lstinline{O2} state are always equal. Randomized abstractions can additionally
express probabilistic relationships, such as the fact that a field of the
\lstinline{O2} state contains a fresh uniformly random value. This is useful
when reasoning about oracles that cache random samples for future use.
Finally, let us mention that randomized state abstraction can be formally shown
to be complete for proving observational equivalence: if \lstinline{O1} and
\lstinline{O2} are observationally equivalent, then this equivalence can be
factored through an intermediate oracle using two randomized state abstractions.
It is formalized in Lean as lemma \lstinline{behavioralRestricted_complete}. 

\subsubsection{Indistinguishability Assumptions}

An indistinguishability proof is parameterized by the assumptions it is allowed
to use. Each assumption is a pair of oracles that are
assumed to be indistinguishable. A proof may use many assumptions,
but for the sake of the computational soundness theorem,
we need to keep track of how many times each assumption is used.
For this reason, we give each assumption an explicit name.
This is modeled by the type \lstinline{IndAssumptions Idx}
that assigns an assumption to each index in \lstinline{Idx}.

Since assumptions may depend on the security parameter, we define
\lstinline{IndAssumptionsFam} to be an indexed set of assumptions
parametrized by the security parameter \(\kappa\). The index type \lstinline{Idx} 
is fixed across all security parameters, but the oracle pair corresponding
to an index may depend on the security parameter \(\kappa\).
For example, the IND-RAND family of assumptions is indexed 
by the singleton type \lstinline{Unit}. 
For each \(\kappa\), the unique index points to
the assumption that \(\mathsf{EncReal}_\kappa\) is indistinguishable from
\(\mathsf{EncRand}_\kappa\).

\subsubsection{Formal Indistinguishability Proofs}

We now describe how the framework represents game-hopping proofs. The central
definition is the type \lstinline{IndistinguishabilityI}. Its inhabitants are
formal proofs certifying that two oracle implementations are indistinguishable
under a given collection of assumptions and, optionally, a bound on the number of
adversarial queries. Before we discuss the constructors of \lstinline{IndistinguishabilityI},
let us start with its declaration:
\begin{center}
\begin{minipage}{\linewidth}
\begin{lstlisting}[
  language=lean,
  numbers=none,
  frame=none,
  aboveskip=0.75\baselineskip
]
inductive IndistinguishableI (A : IndAssumptions Idx) :
    (b : ℕ∞) → (o1 : OracleImpl O) → (o2 : OracleImpl O) → Type
\end{lstlisting}
\end{minipage}
\end{center}

An element of
\lstinline{IndistinguishabilityI A b o1 o2}
is an abstract proof that the oracles \lstinline{o1} and \lstinline{o2} are
indistinguishable, against adversaries that make at most \lstinline{b} queries,
assuming the indistinguishability of all the oracle pairs in \lstinline{A}.
The query bound \lstinline{b} is useful for arguments that first fix the number of oracle queries and then prove security
uniformly for every such bound. We represent bounds using \lstinline{ℕ∞},
corresponding to $\mathbb{N} \cup \{\infty\}$. When no explicit bound is needed,
we use \lstinline{b := ∞}. 

For oracle \emph{families} (indexed by the security parameter), we lift this notion
pointwise. The type \lstinline{Indistinguishable} is used when the query bound is
fixed to \lstinline{∞}: it requires an \lstinline{IndistinguishabilityI} proof
for every security parameter. This is the form used in most of our examples. 
The type \lstinline{IndistinguishableWithQueryBound} keeps the query bound explicit:
it requires
a \lstinline{IndistinguishabilityI} proof for every finite bound \lstinline{b}
and every security parameter \lstinline{κ}.
These two family-level notions are shown in Figure~\ref{fig:indistinguishable}.

\begin{figure}[H]
\centering
\begin{minipage}{\linewidth}
\begin{lstlisting}[
  language=lean,
  aboveskip=0.75\baselineskip
]
def Indistinguishable
    (AFam : IndAssumptionsFam)
    (O1 O2 : (κ : ℕ) → OracleImpl O) : Type :=
  (κ : ℕ) → IndistinguishableI (AFam κ) ∞ (O1 κ) (O2 κ)

def IndistinguishableWithQueryBound
    (AFam : IndAssumptionsFam)
    (O1 O2 : (κ : ℕ) → OracleImpl O) : Type :=
  (b : ℕ) → (κ : ℕ) → IndistinguishableI (AFam κ) b (O1 κ) (O2 κ)
\end{lstlisting}
\end{minipage}
\caption{Lifting indistinguishability to oracle families.}
\label{fig:indistinguishable}
\end{figure}

For example, the type of a proof of IND-CPA security for a certain encryption scheme
that does not use an explicit query can be defined as in Figure~\ref{fig:ind-cpa-proof-type}.

\begin{figure}[H]
\centering
\begin{minipage}{\linewidth}
\begin{lstlisting}[
  language=lean,
  aboveskip=0.75\baselineskip
]
def IndCpaProof (A : IndAssumptionsFam) (s : SymEncSchemeFamily K C) :=
  Indistinguishable A (IndCpaLFam s) (IndCpaRFam s)
\end{lstlisting}
\end{minipage}
\caption{Type of a proof that a certain encryption scheme is IND-CPA.}
\label{fig:ind-cpa-proof-type}
\end{figure}

Finally, let us discuss the constructors of \lstinline{IndistinguishabilityI}.
Each constructor represents either a proof principle for a single
indistinguishability hop, or a structural principle for combining such hops into
larger game-hopping proofs.
\begin{enumerate}[leftmargin=*, itemsep=0.2\baselineskip]
\item \lstinline{IndistinguishabilityI.obsEqB} represents a hop between
      observationally equivalent oracles. Given a query bound \lstinline{b},
      it inputs a proof \lstinline{ObsEqBounded b O1 O2} and outputs
      \lstinline{IndistinguishabilityI A b O1 O2}, for any set of assumptions
      \lstinline{A}. The predicate \lstinline{ObsEqBounded} is the bounded
      version of observational equivalence: it compares the two oracles only on
      query lists of length at most \lstinline{b}. When \lstinline{b := ∞}, it
      coincides with the unbounded notion \lstinline{ObsEq}. Since
      \lstinline{ObsEq} always implies \lstinline{ObsEqBounded}, we
      also provide the derived constructor \lstinline{IndistinguishabilityI.ofObsEq}. It inputs a
      proof \lstinline{ObsEq O1 O2} and outputs
      \lstinline{IndistinguishabilityI A b O1 O2}, for arbitrary
      \lstinline{A} and \lstinline{b}. This derived constructor is the one used
      in most of our examples.

\item \lstinline{IndistinguishabilityI.reduction} represents the hop obtained by
      applying the same oracle reduction on both sides of an indistinguishability
      proof. It inputs a proof
      \lstinline{IndistinguishabilityI A ∞ O1 O2} and outputs
      \lstinline{IndistinguishabilityI A b (R ◇ O1) (R ◇ O2)}. The premise is
      unbounded because \lstinline{R} may ask several queries to the underlying
      oracle while answering a single external query. Thus a bound
      \lstinline{b} on the composed oracle does not, by itself, determine a
      bound on the number of queries made to \lstinline{O1} or \lstinline{O2}.
      One could refine the constructor by also tracking query expansion of
      reductions: for example, if \lstinline{R} asked at most $q$ oracle queries
      per external query, the premise could use the bound $q \cdot b$ instead
      of $\infty$. We do not include this refinement, since we are not aware of
      examples that require this kind of reasoning.

\item \lstinline{IndistinguishabilityI.assumption} represents a hop between
      two oracles that are assumed to be indistinguishable. It inputs an index
      \lstinline{i : Idx} and outputs
      \lstinline{IndistinguishabilityI A b O1 O2}, where
      \lstinline{(O1, O2)} is the pair of oracles stored in \lstinline{A} under
      the index \lstinline{i}.

\item \lstinline{IndistinguishabilityI.symm} represents symmetry of the
      indistinguishability relation. It inputs
      \lstinline{IndistinguishabilityI A b O1 O2} and outputs
      \lstinline{IndistinguishabilityI A b O2 O1}.

\item \lstinline{IndistinguishabilityI.trans} represents transitivity of
      indistinguishability, and is used to compose adjacent hops. It inputs
      proofs \lstinline{IndistinguishabilityI A b O1 O2} and
      \lstinline{IndistinguishabilityI A b O2 O3}, and outputs a proof
      \lstinline{IndistinguishabilityI A b O1 O3}.

\item \lstinline{IndistinguishabilityI.longSequence} represents the
      transitivity principle used in hybrid arguments. It inputs a natural
      number $l$, a sequence of oracles $O_0, O_1, \ldots, O_l$
      and, for every $i < l$, a proof
      \lstinline{IndistinguishabilityI A b (O i) (O (i + 1))}. It outputs a proof
      \lstinline{IndistinguishabilityI A b (O 0) (O l)}. This principle could be
      derived by repeated applications of \lstinline{trans}, but we keep it as
      a separate constructor to distinguish two uses of transitivity. Repeated
      applications of \lstinline{trans} describe a fixed sequence of hops,
      whereas \lstinline{longSequence} describes a family of hybrid arguments
      whose length may depend on parameters such as the security parameter
      $\kappa$ or the query bound \lstinline{b}.
\end{enumerate}

\subsubsection{Automation and Domain-Specific Tactics}

Let us now discuss the tactics introduced by our framework that automate the construction
of \lstinline{IndistinguishabilityI} proofs. The main
tactic is \lstinline{game_hopping}. It takes a fixed list of intermediate
oracles written by the user in the proof script,
\[
  \mathcal{O}_0,\mathcal{O}_1,\ldots,\mathcal{O}_n
\]
whose first and last elements match the current goal, and generates one
subgoal for each concrete adjacent hop
\[
  \mathcal{O}_i \approx \mathcal{O}_{i+1}.
\]
It then tries to solve each subgoal automatically, using two methods:
first, by recognizing a common reduction applied to both sides of an available
assumption; and second, by proving observational equivalence. The resulting hop
proofs are combined by repeated applications of the \lstinline{trans}
constructor. Any adjacent pair that cannot be solved automatically is left as a
subgoal for the user.

The assumption case is syntactic. The tactic tries to recognize goals of the
form
\begin{leandisplay}
\leanline!R ◇ O1 ≈ R ◇ O2!
\end{leandisplay}
where the same reduction is applied to both sides. It then checks whether
\(\mathcal{O}_1\) and \(\mathcal{O}_2\), possibly in the opposite order, match
one of the assumptions available in the current \lstinline{IndAssumptions}.
Since assumptions are stored in an indexed family, the tactic must also
guess the relevant index. This is done by enumerating simple index types,
including units, booleans, products, and disjoint sums. This lightweight search
is sufficient for our examples.

The more interesting case is observational equivalence. For this, the framework
provides tactics
\begin{leandisplay}
\leanline!by_abstraction f!
\qquad
\leanline!by_abstraction ← f!
\end{leandisplay}
The first tactic attempts to prove that \(f\) is a correct deterministic state
abstraction from the left oracle to the right oracle; the second uses the same
abstraction in the opposite direction. There are analogous tactics
\begin{leandisplay}
\leanline!by_rand_abstraction f!
\qquad
\leanline!by_rand_abstraction ← f!
\end{leandisplay}
for randomized state abstractions.
The \lstinline{game_hopping} tactic only tries a small fixed collection of
abstraction candidates. These include the identity map, which is useful when
the two oracles have the same state type, and simple maps that add or remove a
unit component, which occur when composing with a stateless reduction.
If a hop requires a more nuanced abstraction, the automatic attempt fails and
the user can provide the abstraction explicitly using \lstinline{by_abstraction}
or \lstinline{by_rand_abstraction} in the remaining subgoal.

The \lstinline{by_abstraction} tactic, and its variants, reduce the goal to the
local correctness conditions for a state abstraction. This requires proving that
the abstraction preserves the initial-state distribution, and that it is
preserved by every query from every starting state. The second obligation usually
requires more work. The tactic handles it by introducing the starting state and
the query, and splitting on the possible query constructors. It then unfolds
oracle definitions marked with the local attribute
\lstinline{game_hopping_unfold} and applies several groups of simplification
lemmas. The first group, \lstinline{sReduction}, eliminates the
syntactic overhead introduced by oracle reductions. Its purpose is to inline the
composition of a reduction with an oracle, so that an expression of the form
\lstinline{R ◇ O} is displayed as an ordinary \lstinline{RState} oracle. After
this step, the remaining goal should no longer expose the internal plumbing of
reduction application or calls to \lstinline{simulateQ}. The next group,
\lstinline{sStateT}, simplifies the resulting stateful computation. Its goal
is to push the state \lstinline{st} through the \lstinline{RState} monad,
expanding \lstinline{get}, \lstinline{put}, and stateful binds until the goal is
reduced, as far as possible, to an equality of pure probabilistic computations.

At that point the main simplification procedure,
\lstinline{sPMF}, takes over. This tactic normalizes
\lstinline{PMF} expressions by applying algebraic laws for probabilistic binds.
In particular, among other rewrites, it uses simprocs that put independent
samples into a canonical syntactic order, performing the following rewrite:
\begin{center}
\leanline!let a <- X; let b <- Y; cont a b!
\quad \(\leadsto\) \quad
\leanline!let b <- Y; let a <- X; cont a b!
\end{center}
whenever, by inspecting the continuation syntactically, the occurrence of
\lstinline{b} precedes the occurrence of \lstinline{a}. It also removes unused
samples of the form
\begin{center}
\leanline!let a <- X; cont!
\quad \(\leadsto\) \quad
\leanline!cont!
\end{center}
whenever \lstinline{a} does not occur in \lstinline{cont}. These syntactic
normalizations are often enough to discharge the goal automatically.
If these normalizations do not close the goal, the tactic runs
\lstinline{split_ifs}, enabling reasoning by cases on the conditional branches
used by the oracle, followed by \lstinline{simp_all}, which uses the hypotheses
generated by these case splits to try to discharge the remaining goals. Any
goals that remain are left for the user to solve manually. Since
\lstinline{split_ifs} can obscure the structure of the goal when automation
fails, we also provide \emph{basic} variants of the tactics, such as
\lstinline{by_abstraction_basic}. These variants omit the \lstinline{split_ifs}
and \lstinline{simp_all} steps, leaving the goal in a more manageable form.
Finally, the tactic \lstinline{simp only [GameHoppingPrettyPrintPMF]} can be used to
restore the do notation for \lstinline{PMF} expressions, which is often more readable than the
normalized form produced by \lstinline{sPMF}.

\subsubsection{IND-RAND to IND-CPA Proof}

We are now ready to present the formalization of the IND-RAND-to-IND-CPA
argument from Section~\ref{subsec:oracle-indistinguishability-example}. The
formal proof is shown in Figure~\ref{fig:ind-rand-to-ind-cpa-proof}. 
It is constructed using the \lstinline{game_hopping} tactic, which, when
provided with the list of intermediate oracles, automatically discharges all
adjacent hops.

\begin{figure}[H]
\centering
\begin{minipage}{\linewidth}
\begin{lstlisting}[
  language=lean,
  aboveskip=0.75\baselineskip
]
def indCpaRandImpliesIndCpa (schemeFam : SymEncSchemeFamily K C) :
    IndCpaProof (IndCpaRandAssumptionFam schemeFam) schemeFam := by
  intro κ
  let Enc := schemeFam.scheme κ
  game_hopping [
    IndCpaL Enc,
    (IndCpaRand_to_IndCpaL) ◇ (IndCpaRandReal Enc),
    (IndCpaRand_to_IndCpaL) ◇ (IndCpaRandRand Enc),
    (IndCpaRand_to_IndCpaR) ◇ (IndCpaRandRand Enc),
    (IndCpaRand_to_IndCpaR) ◇ (IndCpaRandReal Enc),
    IndCpaR Enc
  ]
\end{lstlisting}
\end{minipage}
\caption{Complete proof that an IND-RAND-secure symmetric encryption scheme is
IND-CPA secure.}
\label{fig:ind-rand-to-ind-cpa-proof}
\end{figure}

\subsection{Computational Soundness}
\label{subsec:computational-soundness}

In this section, we provide the computational interpretation of the formal
\lstinline{IndistinguishabilityI} proof. Specifically, we describe the
computational soundness theorem, which we also formalized in Lean. At a high
level, it states that if the assumptions used in the proof are hard to
distinguish, then so is its conclusion.

Let us begin by recalling the standard notion of computational
indistinguishability. A distinguisher for an oracle specification
\lstinline{O} is a randomized algorithm that interacts with an oracle of that
specification and returns a Boolean. In our formalization, distinguishers are
represented as follows:
\begin{leandisplay}
\leanline!adversaryT O := OracleComp (withPMFSpec O) Bool!
\end{leandisplay}
The operation \lstinline{runDistinguisher d o : PMF Bool}
runs a distinguisher \lstinline{d : adversaryT O} against an oracle
implementation \lstinline{o : OracleImpl O}. It samples the initial state of
\lstinline{o}, simulates the oracle computation using the query implementation
of \lstinline{o}, and returns the resulting distribution over Booleans.
Given two oracle implementations \lstinline{o1} and \lstinline{o2}, the signed
advantage of \lstinline{d}, denoted \lstinline{advantage d o1 o2}, is the
difference between the probabilities that \lstinline{runDistinguisher d o1}
and \lstinline{runDistinguisher d o2} return \lstinline{true}. Usually,
the advantage is defined as the absolute value of this difference,
but we use the signed version because it has convenient algebraic properties: in
particular, swapping the two oracles negates the advantage, and composing two
hops adds the corresponding advantages.
As briefly mentioned in
Section~\ref{subsec:oracle-indistinguishability-example}, computational
indistinguishability is an asymptotic property of oracle families indexed by the
security parameter. Two oracle families \lstinline{o1} and \lstinline{o2} are
computationally indistinguishable if no polynomial-time family of distinguishers
can achieve a non-negligible advantage in distinguishing \lstinline{o1 κ} from
\lstinline{o2 κ}, with both polynomial time and negligibility measured with
respect to the security parameter \(\kappa\).

The definition of computational indistinguishability relies on the notion of
polynomial time. Unfortunately, reasoning about a function's running time is not
directly supported in Lean. Defining polynomial time would first require
choosing a computational model, such as Turing machines or a suitable
\(\lambda\)-calculus, defining its running time, and expressing the relevant
distinguishers and reductions in that model. While technically possible, this is
currently impractical, especially since cryptographic adversaries require both
randomness and oracle interaction. Developing a convenient framework capable of
supporting this kind of reasoning is an important and interesting problem, but
it is outside the scope of this paper --- some foundational work on computation
models is ongoing in the CSLib project~\cite{cslib}.
For this reason, instead of directly formalizing asymptotic computational
indistinguishability, our computational soundness theorem follows a
concrete-security approach.

\subsubsection{Concrete Security}

In the concrete-security approach, the soundness theorem does not directly
assert asymptotic computational indistinguishability. Instead, it gives an
explicit reductionist interpretation of an
\lstinline{IndistinguishabilityI} proof: every distinguisher for the two oracles
related by the proof can be transformed into distinguishers for the assumptions
used in the proof, and the advantage of the original distinguisher is expressed in terms of
the advantages of these assumption distinguishers.

Specifically, let \lstinline{p} be an indistinguishability proof of type
  \lstinline{IndistinguishabilityI A b o1 o2}
and let \lstinline{d : adversaryT O} be a distinguisher for the common
specification of \lstinline{o1} and \lstinline{o2}, making at most
\lstinline{b} queries. The soundness theorem analyzes the proof \lstinline{p}
and, for each assumption used in the proof, constructs an oracle reduction
from the distinguisher \lstinline{d} to a distinguisher for that assumption.
Composing this reduction with \lstinline{d} yields a distinguisher for the
corresponding assumption.

There is one bookkeeping issue. The constructor \lstinline{symm} reverses the
direction of an indistinguishability proof so that an assumption may be used either
in its original direction or in the opposite direction. We track this by
considering directed assumptions: for each assumption \(a = (O_1,O_2) \in A\),
we consider its reversal \(a^R = (O_2,O_1) \in \bar A\). An occurrence of \(a\)
in \lstinline{p} is counted as a use of \(a^R\) if the path from that occurrence
to the root of the proof passes through an odd number of \lstinline{symm}
constructors; otherwise, it is counted as a use of \(a\).

The computational soundness theorem (see theorem \lstinline{computationalSoundness} in the Lean formalization)
constructs a reduction for each
directed assumption \(a \in A \cup \bar A\) that appears at
least once in \lstinline{p}. Let \(\mathcal{R}_a\) be the corresponding reduction,
and let \(n_a\) be the number of times \(a\) appears in \lstinline{p}.
The theorem then expresses the signed advantage 
of \lstinline{d} as the following sum:
\[
  \mathsf{adv}_d(o_1,o_2)
  =
  \sum_{a \in A \cup \bar A}
  n_a \cdot
  \mathsf{adv}_{d \diamond \mathcal{R}_a}(a.1,a.2).
\]
Consequently, the usual unsigned advantage is bounded by
\[
  |\mathsf{adv}_d(o_1,o_2)|
  \leq
  \sum_{a \in A \cup \bar A}
  n_a \cdot
  |\mathsf{adv}_{d \diamond \mathcal{R}_a}(a.1,a.2)|.
\]
In this way, the theorem asserts that if the directed assumptions used in the
proof are hard to distinguish, then the endpoint
oracles \lstinline{o1} and \lstinline{o2} are hard to distinguish as well.

It remains to explain how the reductions \(\mathcal{R}_a\) are constructed. This
matters because these reductions are applied to the endpoint distinguisher
\lstinline{d} before it is run against the corresponding assumption. Thus, the
security guarantee obtained from the theorem depends not only on the number of
occurrences of each assumption, but also on the complexity of the produced
reductions: the more complex the reductions are, the weaker the guarantee.
The reduction \(\mathcal{R}_a\) is constructed as follows. For each occurrence
of \(a\) in the proof, the theorem follows the path from that occurrence to the
root of the proof and composes the reductions appearing along this path. This
produces a collection of reductions, one for each occurrence of \(a\):
\[
  \mathcal{R}_{a,1}, \ldots, \mathcal{R}_{a,n_a},
\]
The final reduction \(\mathcal{R}_a\) is a
randomized parallel composition of this collection: during initialization, it samples an index
\(k \in \{1,\ldots,n_a\}\) uniformly at random and then behaves like
\(\mathcal{R}_{a,k}\) for the rest of the execution.
Thus, \(\mathcal{R}_a\) is obtained by composing reductions that
already appear in the proof, so its complexity can be judged by inspecting the
proof object. This is one reason why \lstinline{IndistinguishabilityI} lives in
\lstinline{Type} rather than \lstinline{Prop}: we do not want proof irrelevance
to erase information about how the proof was constructed. For a way of
examining a proof object, see definition \lstinline{proof_constants_simp} in
\path{Examples/Proofs/IndCpaRandImpliesIndCpa.lean}.

\subsubsection{Asymptotic Computational Soundness Revisited}
\label{subsec:asymptotic-computational-soundness}

We conclude with a discussion of how, under suitable external complexity assumptions, the concrete soundness theorem can be used to recover the usual asymptotic notion of computational indistinguishability. One purpose of this discussion is to motivate future work on formalizing asymptotic computational indistinguishability in Lean.

Let \(\mathcal{O}_1\) and \(\mathcal{O}_2\) be oracle families indexed by the security parameter \(\kappa\), and suppose we are given a family of formal proofs establishing their indistinguishability, depending on \(\kappa\) and on a query bound \(b\). Assume that these proofs use a finite set of assumptions, perform a number of hops that is polynomial in \(\kappa\) and \(b\), and that each reduction \(\mathcal{R}_a\) constructed in the soundness theorem runs in polynomial time. Then the soundness theorem provides a negligible bound on the adversary's advantage.
In more detail, observe that if we fix a family of distinguishers \(d_\kappa\) whose running time is bounded by a polynomial \(t(\kappa)\), then we can instantiate the query bound \(b\) with \(t(\kappa)\), since an adversary cannot make more queries than its running time. Therefore, the coefficients \(n_a\) are polynomially bounded. The negligibility of the advantage of \(d_\kappa\) now follows immediately from the concrete soundness theorem: it bounds the adversary's advantage by a finite sum of negligible functions.

An important question remains: when are the reductions \(\mathcal{R}_a\) polynomial-time? Informally, we argue that this holds provided the proof family satisfies the following complexity side conditions. These conditions are treated as external assumptions and are not formalized in Lean.
\begin{enumerate}
\item The depth of the proof objects is bounded by a constant, independently of \(\kappa\) and \(b\).
\item Every reduction used in the proof objects runs in polynomial time in \(\kappa\), \(b\), and any indices introduced by surrounding \lstinline{longSequence} steps.
\end{enumerate}


The discussion above assumes the non-uniform setting. Indeed, the reductions
\(\mathcal{R}_{a,\kappa}\) produced by the soundness theorem may depend on the
\(\kappa\)-th proof object. To obtain the same conclusion in a uniform setting,
one should additionally require the proof family to be uniform in \(\kappa\)
and \(b\). Intuitively, this means that the parameters \(\kappa\) and \(b\) may
determine the lengths of hybrid sequences and may appear as parameters to
reductions and intermediate oracles, but they should not otherwise change the
structure of the proof object. A weaker formal condition, which is still
sufficient, is that the relevant proof object can be generated from
\(\kappa\) and \(b\) by a polynomial-time procedure.

\section{Case Studies}
\label{sec:case-studies}

In this section, we present several case studies of the framework.
In a formal game-hopping proof, the main work is typically twofold:
choosing suitable reductions and intermediate games, and proving the
observational equivalences used by the hops. Pen-and-paper proofs usually
focus on the former and leave the latter informal. Giving the full details of
the constructions that we formalize would take up too much space, as it would require
specifying the constructions, their security definitions, the reductions used
in the proofs, and the intermediate games. This is also not the focus of the
paper, which is concerned with ways of reasoning about game-hopping proofs in
Lean. Accordingly, for each case study we briefly recall the cryptographic
context and then focus on the proof idioms that arise in the formalization,
especially those used to prove observational equivalence. These idioms mostly
concern observational equivalence, and we often explain them using small
illustrative examples that isolate the relevant reasoning pattern.

\subsection{Encrypt-then-MAC}
\label{subsec:encrypt-then-mac}

We begin our case studies with the Encrypt-then-MAC construction and its proof
of security under chosen-ciphertext attacks. To state this result, we first
define IND-CCA, a security notion stronger than IND-CPA. At a high level, IND-CCA
security is a version of IND-CPA, where the adversary is additionally allowed to
decrypt ciphertexts of their choice, provided that they were not previously returned
by an encryption query. More specifically, it is expressed as the
indistinguishability of two oracles called \textsc{EncDecLeft} and
\textsc{EncDecRight}. Both support two types of queries:
\(\mathsf{encrypt}(m_0,m_1)\) and \(\mathsf{decrypt}(c)\). The state of each
oracle consists of an encryption key and a set of ciphertexts. When
\textsc{EncDecLeft} receives a query \(\mathsf{encrypt}(m_0,m_1)\), it encrypts
\(m_0\) using the encryption key, returns the resulting ciphertext, and adds it
to the set. When it receives a query \(\mathsf{decrypt}(c)\), it returns the
decryption of \(c\), unless \(c\) belongs to the set, in which case it returns
a fixed default message. The oracle \textsc{EncDecRight} is defined
analogously.

Another primitive required for the Encrypt-then-MAC construction is a message
authentication code (MAC). It is a symmetric-key primitive that allows
parties to authenticate messages. It consists of a key-generation algorithm and
a deterministic tagging algorithm that takes a key and a message and
returns a tag. Intuitively, a secure MAC is unforgeable: without knowing the
key, an adversary cannot produce a valid tag for a message.
We formalize this property as the indistinguishability of two oracles,
\textsc{MacReal} and \textsc{MacIdeal}. Both support two types of queries:
\(\mathsf{tag}(m)\) and \(\mathsf{check}(m,t)\), and both maintain internal
state. The state of \textsc{MacReal} consists only of the secret key, whereas
the state of \textsc{MacIdeal} additionally contains a set of message--tag
pairs. Both oracles answer a query \(\mathsf{tag}(m)\) by computing and
returning a tag \(t\) for \(m\). In addition,
\textsc{MacIdeal} records the pair \((m,t)\) in its internal state. The two
oracles differ in how they answer \(\mathsf{check}(m,t)\) queries.
\textsc{MacReal} recomputes the tag of \(m\) under the secret key and checks
whether it is equal to \(t\), whereas \textsc{MacIdeal} checks whether the pair
\((m,t)\) is recorded in its internal state.

The Encrypt-then-MAC construction shows how to combine an IND-CPA-secure encryption scheme with a secure MAC to obtain an IND-CCA-secure encryption scheme. To encrypt a message, it encrypts the
message using the IND-CPA scheme, computes a tag of the ciphertext, and
returns the ciphertext-tag pair. To decrypt a pair \((c,t)\), it checks
whether \(t\) is a valid tag for \(c\); if so, it decrypts \(c\), and otherwise
returns a fixed default message. Intuitively, the construction is
IND-CCA secure because MAC security makes
decryption queries of \textsc{EncDecLeft} and \textsc{EncDecRight} always return the
default message: the query either has an invalid tag, or uses a ciphertext that
was previously returned by the encryption oracle and is therefore on the
do-not-decrypt list. This observation reduces the IND-CCA
security of the construction to the IND-CPA security of the underlying
encryption scheme.

\subsubsection{Formalization}

The HOPSCOTCH proof that the Encrypt-then-MAC construction is IND-CCA secure is
shown in Figure~\ref{fig:encrypt-then-mac-proof}. The proof follows the lines
of \cite[Section~4.5]{evans2024prooffrog}.
We include the proof to illustrate another use of the
\leanline{game_hopping} tactic, but we do not expect the reader to follow all
of its details from the listing alone. For the full proof see
\path{Examples/Proofs/EncryptThenMacIndCca.lean} in the repository.
There are seven hops in the proof. Three follow the
reduction-applied-to-assumption pattern and are solved automatically by the
\leanline{game_hopping} tactic. The remaining four are
observational-equivalence hops, each requiring an explicit abstraction function
relating the states of the adjacent oracle implementations.
Rather than spelling out the full definitions of those abstractions,
we explain how these abstractions behave, occasionally using small illustrative examples.
This lets us present the main proof ideas without introducing the full internal states of the reductions and oracles.

\begin{figure}
\centering
\begin{minipage}{\linewidth}
\begin{lstlisting}[
  language=lean,
  aboveskip=0.75\baselineskip
]
def indCpaAndMacUfImpliesIndCcaEncryptThenMacFam
    (encFam : SymEncSchemeFamily KEnc (fun _ => BitVec))
    (macFam : MACSchemeFamily KMac Tag) :
    IndCcaProof
      (IndCpaAssumptionFam encFam ⊕ MACUFAssumptionFam macFam)
      (encryptThenMacFamily encFam macFam) := by
  intro κ
  let enc := encFam.scheme κ
  let mac := macFam.scheme κ
  game_hopping [
    IndCcaL (encryptThenMac enc mac),
    (EtMFromMACLReduction enc) ◇ (MACUFReal mac),
    (EtMFromMACLReduction enc) ◇ (MACUFIdeal mac),
    (EtMFromIndCpaReduction mac) ◇ (IndCpaL enc),
    (EtMFromIndCpaReduction mac) ◇ (IndCpaR enc),
    (EtMFromMACRReduction enc) ◇ (MACUFIdeal mac),
    (EtMFromMACRReduction enc) ◇ (MACUFReal mac),
    IndCcaR (encryptThenMac enc mac)
  ]
  · by_abstraction ← EtMMacReductionToIndCcaAbstraction
  · by_abstraction ← EtMIndCpaToMacIdealAbstraction
  · by_abstraction EtMIndCpaToMacIdealAbstraction
  · by_abstraction EtMMacReductionToIndCcaAbstraction
\end{lstlisting}
\end{minipage}
\caption{Proof that the encrypt-then-MAC construction is IND-CCA secure.}
\label{fig:encrypt-then-mac-proof}
\end{figure}

\paragraph{Reshuffling abstraction.}
The first abstraction, \lstinline{EtMMacReductionToIndCcaAbstraction}, simply
reshuffles parts of the oracle state. As a toy example, consider two oracles.
The first has state \lstinline!ℕ × Bool!, while the second has state given by a
record with two fields, \lstinline!n : ℕ! and \lstinline!seen : Bool!. Such a
situation can arise when the first oracle is obtained by applying a reduction
with state \lstinline!ℕ! to an oracle with state \lstinline!Bool!, while the
second oracle is written out directly. If the two implementations differ only in
how this state is packaged, then they can be related by the abstraction
\lstinline!fun (x, b) => { n := x, seen := b }!. This abstraction does not hide
or duplicate any information; it merely translates between two equivalent state
representations.

\paragraph{Introducing redundancy.}
The second abstraction, \lstinline{EtMIndCpaToMacIdealAbstraction}, is more
interesting. It captures the fact that two fields in the state of the target
oracle are always equal. As a toy example, consider two oracles implementing a
set of bitvectors. They support two queries, \lstinline!add! and
\lstinline!check!, both taking a bitvector as input. The first oracle implements
this functionality directly: its state is \lstinline!Finset (BitVec n)!,
\lstinline!add! inserts the input into the set, and \lstinline!check! tests
membership in the set. The second oracle stores two copies of the set,
maintaining a set of type \lstinline!Finset (BitVec n) × Finset (BitVec n)!. Its \lstinline!add!
query inserts the input into both sets, while \lstinline!check! tests whether
the input belongs to both of them. The two implementations are observationally
equivalent because the two sets in the second oracle are always kept equal.
This is proved using the abstraction from the direct implementation to the redundant one,
given by \lstinline!fun s => (s, s)!. This example is less artificial than it may seem:
a similar situation arises in the encrypt-then-MAC
proof, where the set of ciphertext-tag pairs returned by the encryption oracle
is maintained both by the idealized MAC functionality and by the do-not-decrypt
list.

\subsection{ElGamal Encryption}
\label{subsec:elgamal}

The ElGamal encryption scheme \cite{elgamal1985public,tsiounis1998security}
is a group-based asymmetric encryption scheme.
Like any public-key encryption scheme, it has two types of keys: a public key
and a secret key. The public key is used to encrypt messages, while the secret
key is used to decrypt them. Since the public key is available to the adversary,
the security definition must require that even an adversary with access to this
key cannot learn information about the encrypted plaintext.
In the indistinguishability formalization, this is captured by adding a public-key query,
\lstinline{getPk}, to the public-key versions of the IND-CPA and IND-RAND
games. This query takes no input and returns the public key.
For ElGamal, we prove a weaker security property, called one-time uniform
ciphertexts. This is a one-shot version of the public-key version of the
IND-RAND game, where the adversary is allowed to make at most one encryption query.
The one-shot restriction is enforced in the oracle state by a Boolean flag recording whether the
encryption query has already been used. Interestingly, for public-key encryption
schemes, this one-time uniform-ciphertext property is enough to derive IND-RAND
security, as we show in the next case study.

To describe the ElGamal encryption scheme, let \(G\) be a cyclic group of order
\(q\), and let \(g\) be a generator of \(G\). The relevant spaces are
\[
  \text{messages: } G, \qquad
  \text{ciphertexts: } G \times G, \qquad
  \text{public keys: } G, \qquad
  \text{secret keys: } \{0,\ldots,q-1\}.
\]
The secret keys are sampled uniformly at random,
whereas the public keys are computed from the secret keys as $\mathsf{pk} = g^{\mathsf{sk}}$.
The encryption algorithm takes a message \(m \in G\), samples \(r\) uniformly at random from \(\{0,\ldots,q-1\}\),
and returns the following ciphertext:
\[
  (g^r, \mathsf{pk}^r \cdot m).
\]
The decryption algorithm takes a ciphertext \((c_1,c_2)\) and returns $c_2 \cdot c_1^{-\mathsf{sk}}$,
which is easily seen to recover the original message.

The one-time uniform-ciphertext property of ElGamal follows from the
decisional Diffie--Hellman (DDH) assumption \cite{Boneh1998DDH}.
This assumption states that the distributions
\[
  (g^a, g^b, g^{ab})
  \qquad\text{and}\qquad
  (g^a, g^b, g^c)
\]
are computationally indistinguishable, where \(a,b,c\) are sampled uniformly at
random from \(\{0,\ldots,q-1\}\).
To express this assumption in our oracle-based framework, we view each of the
two distributions as a stateless oracle. Each oracle supports a single query,
\(\textsf{sample}\), which takes no input and returns a sample from the
corresponding distribution.

The intuition behind the proof is as follows. An ElGamal ciphertext can be
generated by sampling the triple $(g^{\textsf{sk}}, g^r, g^{\textsf{sk} \cdot r})$
and returning $(g^r, g^{\textsf{sk} \cdot r} \cdot m)$.
Since the adversary has access to the public key, they see a triple \((g^{\textsf{sk}}, g^r, g^{\textsf{sk} \cdot r}\cdot m)\). By DDH, this triple is
indistinguishable from \((g^{\textsf{sk}}, g^r, g^c\cdot m)\), where \(c\) is sampled
uniformly at random. Therefore the ciphertext is indistinguishable from
$(g^r, g^c \cdot m)$. Since \(g\) is a generator of \(G\), the element \(g^c\) is uniformly
distributed over \(G\). Multiplying a uniformly sampled group element by a fixed
message \(m\) does not change this distribution. Thus
\((g^r, g^c \cdot m)\) is distributed as \((x,y)\), where \(x\) and \(y\) are
independent uniform elements of \(G\). It follows that the ciphertext is
computationally indistinguishable from the uniform distribution over
\(G \times G\).

\subsubsection{Formalization}
For space reasons, we do not present a Lean figure with the proof of ElGamal.
See \path{Examples/Proofs/DDHImpliesElGamalOTUCPub.lean} for the full proof, which follows the argument of \cite[Section~3.3]{shoup2004sequences}. It consists of seven hops,
one of which is a reduction-applied-to-assumption hop, and the remaining six are observational equivalence hops.
These observational-equivalence hops require explicit abstraction functions and are solved using \leanline{by_abstraction}
and \leanline{by_rand_abstraction}. The proofs are almost automatic, but they require group-specific simplification
lemmas stored under \leanline{GH_group_random_exp} and \leanline{GH_group_norm}.
In addition to reshuffling, the abstraction functions
use two new idioms (they do not use the redundancy idiom). Below, we present them together with descriptions of the simplification lemmas.

\paragraph{Forgetting irrelevant information.}
This idiom is useful when part of the state never affects the oracle's
responses. Consider the following two oracles. The first oracle has state
\lstinline{G × ℕ}. It is initialized by sampling
\(a \leftarrow \{0,\ldots,q-1\}\) uniformly at random and storing the pair
\((g^a,a)\). It supports a single query, \lstinline{getG}, which returns the
first component of the state. The second oracle has state \lstinline{G} and
behaves in the same way, except that it stores only \(g^a\).
To show that these two oracles are observationally equivalent, we can use
the abstraction \lstinline{fun (x, _) => x},
which forgets the irrelevant second component of the state. A similar
situation arises in the ElGamal proof.

\paragraph{Delaying randomness.}
The second idiom used in the ElGamal proof is randomized abstraction. It is
useful when one oracle samples random values eagerly and stores them for later,
while another oracle samples the same values lazily when they are first
needed. Consider two oracles, both supporting a single query, \lstinline{getN}.
The eager oracle has state \lstinline{ℕ}. It is initialized by sampling a
value uniformly from \(\{0,\ldots,q-1\}\), and answers \lstinline{getN} by
returning the value stored in its state. The lazy oracle has state
\lstinline{Option ℕ} (i.e. it either carries a natural number or is empty).
It is initialized to the empty value \lstinline{none}. When it receives
a \lstinline{getN} query, it checks its state: if the state is empty, it samples
a value uniformly from \(\{0,\ldots,q-1\}\), stores it, and returns it;
otherwise, it returns the value already stored in the state.
To show that these two oracles are observationally equivalent, we use the
following randomized abstraction from the lazy state to the eager one:
\[
  s \mapsto
  \begin{cases}
    \text{uniform over } \{0,\ldots,q-1\}
      & \text{if } s = \mathsf{none},\\
    n
      & \text{if } s = \mathsf{some}(n).
  \end{cases}
\]
This is a valid abstraction because it maps the initial lazy state,
\(\mathsf{none}\), to the uniform initial distribution of the eager oracle. It
also commutes with the \lstinline{getN} query. If the lazy state is
\(\mathsf{none}\), then after the query the lazy oracle produces the
distribution over output-state pairs \((n,\mathsf{some}(n))\), where \(n\) is
sampled uniformly from \(\{0,\ldots,q-1\}\). Applying the abstraction to the
state component maps this to the distribution \((n,n)\), again with \(n\)
uniform. On the other hand, if we apply the abstraction first to the empty state
of the lazy oracle, we obtain the uniform distribution over eager states. The
\lstinline{getN} query then simply copies the state value to the output, giving
the same distribution \((n,n)\) for uniformly sampled \(n\). Thus, replying to a
query commutes with applying the abstraction.

An example like this arises naturally in the ElGamal proof. To use the DDH assumption, we
need to sample the value \(g^r\) together with \(g^{\textsf{sk}}\) during
initialization. In contrast, the original ElGamal oracle samples \(g^r\) only
when answering the encryption query.

\paragraph{Group-specific simplifications.}
To facilitate the formalization of the ElGamal proof (and hopefully future proofs),
we developed two groups of simp lemmas: \lstinline{GH_group_norm} and
\lstinline{GH_group_random_exp}. The first one, \lstinline{GH_group_norm},
collects general algebraic and probabilistic facts about groups. It contains
standard group identities, such as \((g^a)^b = g^{ab}\), most of which are
already present in Mathlib. It also contains probabilistic simplifications
specific to uniform distributions over groups. For example, multiplying a
uniformly sampled group element by a fixed group element preserves the uniform
distribution, which gives rewrites of the following form:
\begin{center}
\leanline!let x <- sampleGroupElem; cont (m * x)!
\quad \(\leadsto\) \quad
\leanline!let x <- sampleGroupElem; cont x!
\end{center}
Here, \lstinline{m} is a fixed group element.

The second attribute, \lstinline{GH_group_random_exp}, contains a single
special-purpose rewrite rule. It rewrites a uniformly sampled exponent applied
to a generator into a uniformly sampled group element:
\begin{center}
\leanline!let x <- sampleExponent; cont (g ^ x)!
\quad \(\leadsto\) \quad
\leanline!let x <- sampleGroupElem; cont x!
\end{center}
We keep this rule separate from \lstinline{GH_group_norm}, because it can fire
in surprising places and produce less convenient goals.

\subsection{One-Time Secrecy Implies IND-CPA}
\label{subsec:ots-implies-ind-cpa}

Our next proof shows that one-time secrecy (OTS), i.e. the one-shot version of
public-key IND-CPA security, implies public-key IND-CPA security.  The
public-key IND-CPA game is analogous to the symmetric-key version: the adversary
submits pairs of messages and receives encryptions of either the left or the
right component.  The only additional feature is that the adversary may also
request the public key through a \lstinline{getPk} query, which takes no input.

The OTS games have the same interface, but, as in the one-time
uniform-ciphertexts game, allow only one non-default encryption query.  Their
state consists of the public key together with a Boolean flag recording whether
an \lstinline{eavesdrop} query has already been answered.  On the first query
\lstinline{eavesdrop(m₀,m₁)}, the left OTS oracle returns an encryption of
\(m_0\), while the right OTS oracle returns an encryption of \(m_1\).
Subsequent \lstinline{eavesdrop} queries return a fixed default ciphertext.
The \lstinline{getPk} query is unrestricted and always returns the public key.

The proof is a hybrid (i.e. \lstinline{longSequence}) argument that explicitly
uses a bound on the number of queries made by the adversary.  Fix a bound \(q\).
We define hybrid oracles
\(\mathcal{H}_i\), for \(0 \leq i \leq q\), where \(\mathcal{H}_i\) answers the
first \(i\) \lstinline{eavesdrop(m₀,m₁)} queries by encrypting \(m_1\), and
all later \lstinline{eavesdrop} queries by encrypting \(m_0\).  Thus
\(\mathcal{H}_0\) is behaviorally equivalent to the left public-key IND-CPA
oracle.  Moreover, because an adversary making at most \(q\) queries cannot see
more than the first \(q\) \lstinline{eavesdrop} queries, \(\mathcal{H}_q\) is
behaviorally equivalent to the right public-key IND-CPA oracle.
It remains to show that adjacent hybrids are indistinguishable.  For each
\(i\), we define a reduction \(\mathcal{R}_i\) that uses an OTS oracle as its
underlying oracle.  The reduction keeps a counter for \lstinline{eavesdrop}
queries.  Before the counter reaches \(i\), it answers queries itself by
requesting the public key from the OTS oracle and encrypting \(m_1\).  When the
counter is exactly \(i\), it forwards the query to the OTS oracle.  After that,
it again answers queries itself, now encrypting \(m_0\).  Therefore
\(\mathcal{R}_i \diamond \mathsf{OTSLeft}\) is behaviorally equivalent to
\(\mathcal{H}_i\), while
\(\mathcal{R}_i \diamond \mathsf{OTSRight}\) is behaviorally equivalent to
\(\mathcal{H}_{i+1}\).  It follows that we can hop from
$\mathcal{H}_i$ to $\mathcal{H}_{i+1}$.
Chaining these \(q\) hops gives indistinguishability of
\(\mathcal{H}_0\) and \(\mathcal{H}_q\), which finishes the proof.

\subsubsection{Formalization}

As in the previous section, we do not include the Lean listing for the proof of
OTS implies IND-CPA. See
\path{Examples/Proofs/OneTimeSecrecyImpliesIndCPAPub.lean}
for the full proof, which follows the outline of \cite[Section~4.4]{evans2024prooffrog}.
The top-level proof consists of three hops: from the left
IND-CPA oracle to \(\mathcal{H}_0\), then from \(\mathcal{H}_0\) to
\(\mathcal{H}_q\), and finally from \(\mathcal{H}_q\) to the right IND-CPA
oracle.  The first hop is an observational-equivalence hop, the second is a
long-sequence hop, and the last is a bounded observational-equivalence hop,
using \lstinline{ObsEqB}.  The long-sequence hop is itself built from the
adjacent hybrid steps.  For each
\(i\), the proof of indistinguishability between \(\mathcal{H}_i\) and
\(\mathcal{H}_{i+1}\) consists of three hops:
from \(\mathcal{H}_i\) to
\(\mathcal{R}_i \diamond \mathsf{OTSLeft}\), then to
\(\mathcal{R}_i \diamond \mathsf{OTSRight}\), and finally to
\(\mathcal{H}_{i+1}\).  The first and last of these are
observational-equivalence hops, while the middle hop is a
reduction-applied-to-assumption hop.

The proof uses all of the deterministic abstraction idioms introduced so far:
reshuffling, introducing redundancy, and forgetting irrelevant information.  It
also uses a specialized bounded abstraction to prove the final bounded
observational equivalence.  The redundancy-introduction and forgetting idioms
appear in slightly different forms than in the previous proof, so we discuss
them below, together with the bounded observational-equivalence abstraction.

\paragraph{Forgetting irrelevant information.}
Consider the first hop from the left IND-CPA oracle to \(\mathcal{H}_0\).
The state of the left IND-CPA oracle consists only of the public key, sampled
during initialization.  The state of the hybrid oracle \(\mathcal{H}_0\)
consists of a counter for the number of \lstinline{eavesdrop} queries and the
public key.  The counter is read and incremented, but it is irrelevant to the
oracle's responses: \(\mathcal{H}_0\) always encrypts the left message.
Therefore, to prove observational equivalence between the two oracles, we
abstract the counter away, using the abstraction
\lstinline{fun (_, pk) => pk}.

\paragraph{Introducing redundancy.}
Now consider the hop from \(\mathcal{H}_i\) to
\(\mathcal{R}_i \diamond \mathsf{OTSLeft}\).  The state of \(\mathcal{H}_i\)
consists of a counter for the number of \lstinline{eavesdrop} queries and the
public key.  The state of
\(\mathcal{R}_i \diamond \mathsf{OTSLeft}\) consists of the reduction's counter
together with the state of the OTS oracle, namely the public key and a flag
recording whether the OTS oracle has already answered its one non-default
\lstinline{eavesdrop} query.  However, the value of this flag is determined by
the counter: it is true exactly when the counter is greater than \(i\).  For
this reason, we prove the equivalence using the abstraction
\lstinline{fun (c, pk) => (c, pk, c > i)}.

\paragraph{Bounded observational equivalence.}
Finally, consider the hop from \(\mathcal H_q\) to the right IND-CPA oracle.
We want to show that these two oracles are indistinguishable for adversaries
making at most \(q\) queries.
To capture this, we use a bounded version of abstraction.  A bounded
abstraction between oracles \(\mathcal O_1\) and \(\mathcal O_2\) consists of
the usual abstraction function \(f\), mapping states of \(\mathcal O_1\) to
states of \(\mathcal O_2\), together with a valuation function \(v\), mapping
states of \(\mathcal O_1\) to \(\mathbb N\).  The valuation of every possible
initial state of \(\mathcal O_1\) must be at least the query bound \(q\), and
processing a query should not decrease the valuation by more than one.  The
usual commutation condition is required only for starting states with non-zero
valuation.  The proof that such a bounded abstraction implies bounded
observational equivalence follows the same structure as the ordinary
abstraction argument; see
\lstinline{correctAbstractionBoundImpliesObsEqBounded} in the formalization.

We apply this technique to the hop between \(\mathcal H_q\) and the right
IND-CPA oracle.  The state of \(\mathcal H_q\) consists of a counter for the
number of \lstinline{eavesdrop} queries and the public key, while the state of
the right IND-CPA oracle consists only of the public key.  The abstraction
function \lstinline{fun (_, pk) => pk} forgets the counter, and the valuation
function \lstinline{fun (c, _) => q - c} counts how many
\lstinline{eavesdrop} queries remain before the bound is reached.
This is a correct bounded abstraction, but proving it in Lean might
be cumbersome, as the definition of the bounded abstraction uses
reasoning about \lstinline{PMF.support}.
For this reason, the formal proof uses a simpler indexed variant,
\lstinline{correctAbstractionB}. It is less general, but easier to use in
practice.

\subsection{GGM: Lifting Pseudorandom Generators to Pseudorandom Functions}
\label{subsec:ggm}

Our final case study is the GGM construction~\cite{goldreich1986construct},
which transforms a length-doubling pseudorandom generator into a
pseudorandom function.  The formal proof is substantially larger than the
previous case studies, so we describe only its main structure.  It illustrates
two features of the framework: a hybrid sequence of length \(\kappa\), and the
delayed-randomness idiom from the ElGamal proof, now applied to entire caches
of random labels rather than to a single sampled value.  These two features
appear essential for handling arbitrary-depth instances of the construction.
To the best of our knowledge, this is the first formal proof of the GGM
construction at nonconstant depth and the first in a general-purpose proof
assistant; CryptoZoo~\cite{brzuska2024cryptozoo} formalizes the construction
at depth~\(3\).

We begin by recalling the underlying primitive.
A length-doubling PRG with seed length \(\kappa\) is a deterministic function
\[
  \mathsf{prg} : \textsf{BitVec }\kappa \to \textsf{BitVec }(2\kappa).
\]
It is secure if the distribution obtained by sampling a uniform seed
\(s \leftarrow \textsf{BitVec }\kappa\) and returning \(\mathsf{prg}(s)\) is
indistinguishable from the uniform distribution over
\(\textsf{BitVec }(2\kappa)\). As usual, we express this as the
indistinguishability of two stateless oracles, both supporting a single query
\lstinline{sample}. The real oracle samples a fresh seed \(s\) and returns
\(\mathsf{prg}(s)\), while the random oracle returns a uniformly sampled element
of \(\textsf{BitVec }(2\kappa)\).

The resulting primitive is a PRF. In our formalization, the input length of
the PRF is also \(\kappa\), so the same parameter controls both the seed length
and the number of PRG iterations. Thus the PRF is a deterministic function of
type
\[
  \mathsf{prf} :
    \textsf{BitVec }\kappa \times
    \textsf{BitVec }\kappa
    \to \textsf{BitVec }\kappa .
\]
It is secure if, for a uniformly sampled seed \(s\), the function
\(x \mapsto \mathsf{prf}(s,x)\) is indistinguishable from a uniformly sampled
function
\[
  r : \textsf{BitVec }\kappa \to \textsf{BitVec }\kappa .
\]
We again express this as indistinguishability of two oracles supporting a
single query, \lstinline{eval(x)}. The real oracle samples a seed \(s\) during
initialization and answers a query \(x\) with \(\mathsf{prf}(s,x)\). The ideal
oracle samples a total function \(r\) during initialization and answers \(x\)
with \(r(x)\).

The GGM construction works as follows. Given a length-doubling PRG
\(\mathsf{prg}\), we define \(\mathsf{prg}_0(s)\) and
\(\mathsf{prg}_1(s)\) to be the first and second halves of
\(\mathsf{prg}(s)\), respectively. Then the GGM pseudorandom function is
defined by
\[
  \mathsf{prf}(s,x)
  =
  \mathsf{prg}_{x_\kappa}
  \bigl(
    \mathsf{prg}_{x_{\kappa-1}}
    (\ldots
      \mathsf{prg}_{x_1}(s)
    \ldots)
  \bigr),
\]
where \(x_i\) is the \(i\)-th bit of the input \(x\), in the order used by the
implementation.

The security of GGM is proved using a hybrid argument, formalized using
\lstinline{longSequence} with \(\kappa\) steps. The \(i\)-th hybrid
\(\mathcal H_i\) maintains a random function
\[
  f_i : \textsf{BitVec }i \to \textsf{BitVec }\kappa,
\]
initialized uniformly at random. On a query
\(x=(x_1,\ldots,x_\kappa) \in \textsf{BitVec }\kappa\), the oracle returns
\[
  \mathcal H_i(x)
  =
  \mathsf{prg}_{x_\kappa}
  \bigl(
    \mathsf{prg}_{x_{\kappa-1}}
    (\ldots
      \mathsf{prg}_{x_{i+1}}(f_i(x_1,\ldots,x_i))
    \ldots)
  \bigr).
\]
The endpoints are the target oracles. Since \(\textsf{BitVec }0\) is a
singleton, sampling
\[
  f_0 : \textsf{BitVec }0 \to \textsf{BitVec }\kappa
\]
is equivalent to sampling a single seed. Thus \(\mathcal H_0\) is the real GGM
oracle. At the other endpoint, no PRG applications remain, so
\(\mathcal H_\kappa\) always answers according to a random function
\(f_\kappa\).
Thus \(\mathcal H_\kappa\) is the ideal PRF oracle.

The central step is to prove that each pair of adjacent hybrids,
\(\mathcal H_i\) and \(\mathcal H_{i+1}\), is indistinguishable.
Rather than give the details of this step, we describe the two patterns that
arise in the proof of GGM security.

\subsubsection{Formalization}

Our formalization, together with some explanatory comments, is available in
\path{Examples/Proofs/GGMSecurePRF.lean}. Here we explain the two main
patterns of the proof, both of which are instances of the delayed-randomness
idiom.

\paragraph{Lazy cache.}
Consider an alternative, lazy implementation of the ideal PRF oracle. Instead
of storing the entire random function in its state, it stores a partial function
\[
  r : \mathsf{BitVec}\ \kappa \to \mathsf{BitVec}\ \kappa \cup \{\bot\} .
\]
When the oracle receives a query \(\textsf{eval}(x)\), it checks whether \(x\)
is already in the domain of \(r\). If so, it returns \(r(x)\). Otherwise, it
samples a fresh value \(y\) uniformly at random from
\(\mathsf{BitVec}\ \kappa\), updates the cache by setting \(r(x) := y\), and
returns \(y\). To prove that this lazy implementation is equivalent to the
eager one (i.e., the ideal PRF oracle), we use a randomized abstraction. The
abstraction maps a partial function \(r\) to a distribution
over total functions by keeping the values already stored in \(r\) fixed, and
sampling independent uniform values for all inputs outside the domain of \(r\).

The main advantage of the lazy implementation is efficiency: it samples values
only for inputs that are actually queried. Replacing the eager implementation
with a lazy one therefore allows us to use polynomial-time reductions, as
required by the computational definition of indistinguishability. The
definition does not, however, require either the endpoint oracles or the
intermediate oracles to run in polynomial time.

\paragraph{Changing the granularity of the lazy cache.}
The second use of delayed randomness changes the granularity at which the lazy
cache is filled. Consider a slightly more eager lazy cache: when queried on
\(x\), it samples fresh random values not only for \(x\), but also for the input
obtained by flipping the last bit of \(x\). Thus the cache is filled in pairs of
inputs sharing the same prefix. We prove that this oracle is equivalent to the
ordinary lazy cache by transitivity through the eager random-function oracle.
Both lazy implementations are related
to the eager oracle by the same randomized abstraction as above: complete the
partial cache by sampling all missing values uniformly at random.

This paired-cache oracle appears naturally in the GGM proof, since a single PRG
call produces two outputs, \(\mathsf{prg}_0(s)\) and \(\mathsf{prg}_1(s)\).


\section{Conclusion and Future Work}
\label{sec:conclusions}
There are several natural directions for extending HOPSCOTCH. First, we plan to formalize additional cryptographic constructions and protocols, using these case studies both to evaluate the framework and to identify recurring proof patterns that can be incorporated into stronger tactics. Second, the explicit structure of HOPSCOTCH proof objects makes the framework a promising target for machine-assisted proof generation. In particular, we would like to investigate whether large language models can construct formal game-hopping proofs while preserving the structure and readability of their pen-and-paper counterparts. Third, we plan to extend the proof calculus with error-bounded transitions, including reasoning based on bad events, possibly by integrating techniques from probabilistic relational Hoare logic. Finally, formalizing probabilistic polynomial time would remove the remaining externally checked complexity conditions and yield fully mechanized computational-security statements.

\begin{acks}
This work was partially supported by the European Research Council (ERC) under
the European Union's Horizon 2020 Research and Innovation Programme (Grant
No.~PROCONTRA-885666), and partially supported by the U.S. National Science
Foundation (NSF) under Award No.~2411704.
\end{acks}

\FloatBarrier
\pagebreak
\bibliographystyle{ACM-Reference-Format}
\bibliography{references}

@InProceedings{Moura2021,
  author    = {Leonardo de Moura and Sebastian Ullrich},
  booktitle = {Automated Deduction - {CADE} 28 - 28th International Conference on Automated Deduction, Virtual Event, July 12-15, 2021, Proceedings},
  title     = {The Lean 4 Theorem Prover and Programming Language},
  year      = {2021},
  editor    = {Andr{\'{e}} Platzer and Geoff Sutcliffe},
  pages     = {625--635},
  publisher = {Springer},
  series    = {Lecture Notes in Computer Science},
  volume    = {12699},
  bibsource = {dblp computer science bibliography, https://dblp.org},
  doi       = {10.1007/978-3-030-79876-5_37},
}

@misc{cslib,
  title={CSLib: The Lean Computer Science Library}, 
  author={Clark Barrett and Swarat Chaudhuri and Fabrizio Montesi and Jim Grundy and Pushmeet Kohli and Leonardo de Moura and Alexandre Rademaker and Sorrachai Yingchareonthawornchai},
  year={2026},
  eprint={2602.04846},
  archivePrefix={arXiv},
  primaryClass={cs.LO},
  url={https://arxiv.org/abs/2602.04846}, 
}

@phdthesis{CertiCrypt,
  TITLE = {{Formal certification of game-based cryptographic proofs}},
  AUTHOR = {Zanella-B{\'e}guelin, Santiago},
  URL = {https://pastel.hal.science/pastel-00584350},
  NUMBER = {2010ENMP0050},
  SCHOOL = {{{\'E}cole Nationale Sup{\'e}rieure des Mines de Paris}},
  YEAR = {2010},
  MONTH = Dec,
  TYPE = {Theses},
  HAL_ID = {pastel-00584350},
  HAL_VERSION = {v1},
}

@misc{fcf,
      title={The Foundational Cryptography Framework}, 
      author={Adam Petcher and Greg Morrisett},
      year={2014},
      eprint={1410.3735},
      archivePrefix={arXiv},
      primaryClass={cs.PL},
      url={https://arxiv.org/abs/1410.3735}, 
}

@article{CryptHOL,
  title={CryptHOL: Game-Based Proofs in Higher-Order Logic},
  author={David A. Basin and Andreas Lochbihler and S. Reza Sefidgar},
  journal={Journal of Cryptology},
  year={2020},
  volume={33},
  pages={494 - 566},
  url={https://api.semanticscholar.org/CorpusID:8723421}
}

@INPROCEEDINGS{SSProve,
  author={Abate, Carmine and Haselwarter, Philipp G. and Rivas, Exequiel and Muylder, Antoine Van and Winterhalter, Théo and Hriţcu, Cătălin and Maillard, Kenji and Spitters, Bas},
  booktitle={2021 IEEE 34th Computer Security Foundations Symposium (CSF)}, 
  title={SSProve: A Foundational Framework for Modular Cryptographic Proofs in Coq}, 
  year={2021},
  volume={},
  number={},
  pages={1-15},
  doi={10.1109/CSF51468.2021.00048}}

@misc{VCVio,
      author = {Devon Tuma and Quang Dao and James Waters and Alexander Hicks and Nicholas Hopper},
      title = {{VCVio}: Verified Cryptography in Lean via Oracle Effects and Handlers},
      howpublished = {Cryptology {ePrint} Archive, Paper 2026/899},
      year = {2026},
      url = {https://eprint.iacr.org/2026/899}
}

@inproceedings{EasyCrypt,
  title={Computer-aided security proofs for the working cryptographer},
  author={Barthe, Gilles and Gr{\'e}goire, Benjamin and Heraud, Sylvain and B{\'e}guelin, Santiago Zanella},
  booktitle={Annual Cryptology Conference},
  pages={71--90},
  year={2011},
  organization={Springer}
}

@article{blanchet2008computationally,
  title={A computationally sound mechanized prover for security protocols},
  author={Blanchet, Bruno},
  journal={IEEE Transactions on Dependable and Secure Computing},
  volume={5},
  number={4},
  pages={193--207},
  year={2008},
  publisher={IEEE}
}

@article{metere2023direct,
  title={A Direct Lazy Sampling Proof Technique in Probabilistic Relational Hoare Logic},
  author={Metere, Roberto and Dong, Changyu},
  journal={arXiv preprint arXiv:2311.16844},
  year={2023}
}

@article{shoup2004sequences,
  title={Sequences of games: a tool for taming complexity in security proofs},
  author={Shoup, Victor},
  journal={cryptology eprint archive},
  year={2004}
}

@book{rosulek2021joy,
  title={The joy of cryptography},
  author={Rosulek, Mike},
  year={2021},
  publisher={Oregon State University}
}

@article{evans2024prooffrog,
  title={ProofFrog: A tool for verifying game-hopping proofs},
  author={Evans, Ross},
  year={2024},
  publisher={University of Waterloo}
}

@article{goldreich1986construct,
  title={How to construct random functions},
  author={Goldreich, Oded and Goldwasser, Shafi and Micali, Silvio},
  journal={Journal of the ACM (JACM)},
  volume={33},
  number={4},
  pages={792--807},
  year={1986},
  publisher={ACM New York, NY, USA}
}

@inproceedings{tsiounis1998security,
  title={On the security of ElGamal based encryption},
  author={Tsiounis, Yiannis and Yung, Moti},
  booktitle={International Workshop on Public Key Cryptography},
  pages={117--134},
  year={1998},
  organization={Springer}
}

@article{elgamal1985public,
  title={A public key cryptosystem and a signature scheme based on discrete logarithms},
  author={ElGamal, Taher},
  journal={IEEE transactions on information theory},
  volume={31},
  number={4},
  pages={469--472},
  year={1985},
  publisher={IEEE}
}

@inproceedings{Boneh1998DDH,
  author    = {Dan Boneh},
  title     = {The Decision Diffie--Hellman Problem},
  booktitle = {Algorithmic Number Theory},
  series    = {Lecture Notes in Computer Science},
  volume    = {1423},
  pages     = {48--63},
  publisher = {Springer},
  year      = {1998}
}

@inproceedings{brzuska2024cryptozoo,
  title={CryptoZoo: A Viewer for Reduction Proofs},
  author={Brzuska, Chris and Egger, Christoph and Puniamurthy, Kirthivaasan},
  booktitle={International Conference on Applied Cryptography and Network Security},
  pages={3--25},
  year={2024},
  organization={Springer}
}

@misc{achim2025aristotleimolevelautomatedtheorem,
  title         = {Aristotle: IMO-level Automated Theorem Proving},
  author        = {Tudor Achim and Alex Best and Alberto Bietti and Kevin Der
                   and Mathïs Fédérico and Sergei Gukov
                   and Daniel Halpern-Leistner and Kirsten Henningsgard
                   and Yury Kudryashov and Alexander Meiburg
                   and Martin Michelsen and Riley Patterson
                   and Eric Rodriguez and Laura Scharff
                   and Vikram Shanker and Vladmir Sicca
                   and Hari Sowrirajan and Aidan Swope and Matyas Tamas
                   and Vlad Tenev and Jonathan Thomm
                   and Harold Williams and Lawrence Wu},
  year          = {2025},
  eprint        = {2510.01346},
  archivePrefix = {arXiv},
  primaryClass  = {cs.AI},
  url           = {https://arxiv.org/abs/2510.01346}
}

@article{singh2025openai,
  title={Openai gpt-5 system card},
  author={Singh, Aaditya and Fry, Adam and Perelman, Adam and Tart, Adam and Ganesh, Adi and El-Kishky, Ahmed and McLaughlin, Aidan and Low, Aiden and Ostrow, AJ and Ananthram, Akhila and others},
  journal={arXiv preprint arXiv:2601.03267},
  year={2025}
}

@misc{deepseekV4,
      title={DeepSeek-V4: Towards Highly Efficient Million-Token Context Intelligence}, 
      author={DeepSeek-AI and Anyi Xu and Bangcai Lin and Bing Xue and Bingxuan Wang and Bingzheng Xu and Bochao Wu and Bowei Zhang and Chaofan Lin and Chen Dong and Chenchen Ling et al},
      year={2026},
      eprint={2606.19348},
      archivePrefix={arXiv},
      primaryClass={cs.CL},
      url={https://arxiv.org/abs/2606.19348}, 
}

@article{vcvioOld,
  title={VCVio: A Formally Verified Forking Lemma and Fiat-Shamir Transform, via a Flexible and Expressive Oracle Representation},
  author={Tuma, Devon and Hopper, Nicholas},
  journal={Cryptology ePrint Archive},
  year={2024}
}

\end{document}